\documentclass[twocolumn]{aastex63}

\hypersetup{linkcolor=blue,citecolor=blue,filecolor=blue,urlcolor=blue}

\newcommand{\sub}[1]{_{\rm #1}}

 \newcommand{\brafracket}[2]{\left( \frac{#1}{#2} \right)}

 \newcommand{\braa}{\left[}
 \newcommand{\kett}{\right]}

\newcommand{\tdisk}{t\sub{disk}}

\newcommand{\SigdotPE}{\dot{\Sigma}_{\rm PE}}
\newcommand{\SigdotX}{\dot{\Sigma}_{\rm X}}
\newcommand{\SigdotEUV}{\dot{\Sigma}_{\rm EUV}}
\newcommand{\SigdotFUV}{\dot{\Sigma}_{\rm FUV}}
\newcommand{\SigdotoX}{\dot{\Sigma}_{\rm X,0}}
\newcommand{\Sigdoto}{\dot{\Sigma}_{\rm FUV,0}}

\newcommand{\cs}{c\sub{s}}
\newcommand{\Mdot}{\dot{M}}
\newcommand{\Mdotacc}{\Mdot\sub{acc}}
\newcommand{\MdotPE}{\Mdot\sub{PE}}
\newcommand{\MdotX}{\Mdot\sub{X}}
\newcommand{\MdotFUV}{\Mdot\sub{FUV}}
\newcommand{\MdotEUV}{\Mdot\sub{EUV}}

\newcommand{\Macc}{M\sub{acc}}
\newcommand{\MPE}{M\sub{PE}}

\newcommand{\Mstar}{M_\star}
\newcommand{\Lstar}{L_\star}
\newcommand{\Rstar}{R_\star}
\newcommand{\Teff}{T\sub{eff}}
\newcommand{\Prot}{P\sub{rot}}
\newcommand{\Ro}{\mathrm{Ro}}
\newcommand{\LX}{L\sub{X}}
\newcommand{\RX}{R\sub{X}}
\newcommand{\PhiEUV}{ \Phi\sub{EUV} } 
\newcommand{\PhiEUVph}{ \Phi\sub{EUV,ph} }
\newcommand{\PhiEUVcor}{ \Phi\sub{EUV,cor} } 
\newcommand{\LFUV}{L\sub{FUV}}
\newcommand{\LFUVacc}{L\sub{FUV,acc}}
\newcommand{\LFUVph}{L\sub{FUV,ph}}
\newcommand{\LFUVcor}{L\sub{FUV,chr}}
\newcommand{\fredEUV}{f\sub{EUV} }
\newcommand{\fredFUV}{f\sub{FUV} }
\newcommand{\Msun}{\rm M_{\sun}}

\newcommand{\Rsun}{\rm R_{\sun}}
\newcommand{\Lsun}{\rm L_{\sun}}
\newcommand{\gsun}{\rm g_{\sun}}
\newcommand{\amu}{\mathrm{m_u}}
\newcommand{\SB}{\sigma\sub{SB}}
\newcommand{\kB}{\mathrm{k_B}}
\newcommand{\au}{{\rm au}}
\newcommand{\yr}{{\rm yr}}

\newcommand{\K}{{\rm K}}

\newcommand{\Tmid}{T\sub{mid}}
\newcommand{\Md}{M\sub{disk}}
\newcommand{\Mdini}{M\sub{d,ini}}

\newcommand{\tauPE}{\tau\sub{PE}}
\newcommand{\tvis}{ \tau\sub{vis} }
\newcommand{\tKH}{ \tau\sub{KH} }

\newcommand{\Rosat}{ {\rm Ro_{sat}} }
\newcommand{\Rofloor}{ {\rm Ro_{floor}} }

\usepackage{bm}

\received{2020 April 24}
\revised{2021 January 9}
\accepted{2021 January 11}
\published{2021 March 10}
\shorttitle{Disk Dispersal around Evolving Intermediate-Mass Stars}
\shortauthors{Kunitomo et al.}
\graphicspath{{./}{fig/}}

\begin{document}

\title{Photoevaporative Dispersal of Protoplanetary Disks around Evolving Intermediate-mass Stars}

\correspondingauthor{Masanobu Kunitomo}
\email{kunitomo.masanobu@gmail.com}

\author[0000-0002-1932-3358]{Masanobu Kunitomo}
\affiliation{Department of Physics, School of Medicine, Kurume University, 67 Asahimachi, Kurume, Fukuoka 830-0011, Japan}

\author[0000-0002-9676-3891]{Shigeru Ida}
\affiliation{Earth-Life Science Institute, Tokyo Institute of Technology, 2-12-1 Ookayama, Meguro-ku, Tokyo 152-8550, Japan}

\author{Taku Takeuchi}
\altaffiliation{Present affiliation: Sanoh Industrial Co., Ltd., Japan}
\affiliation{Department of Earth and Planetary Sciences, 
Tokyo Institute of Technology, 2-12-1 Ookayama, Meguro-ku, 
Tokyo 152-8551, Japan}

\author[0000-0002-6648-2968]{Olja Pani\'c}
\altaffiliation{Royal Society Dorothy Hodgkin Fellow}
\affiliation{School of Physics and Astronomy, University of Leeds, Leeds LS2 9JT, UK}

\author[0000-0002-1575-680X]{James M. Miley}
\affiliation{School of Physics and Astronomy, University of Leeds, Leeds LS2 9JT, UK}
\affiliation{Joint ALMA Observatory, Alonso de Cordova 3107, Vitacura, Santiago, Chile }
\affiliation{National Astronomical Observatory of Japan, Alonso de Cordova 3788, 61B Vitacura, Santiago, Chile}

\author[0000-0001-9734-9601]{Takeru K. Suzuki}
\affiliation{School of Arts \& Sciences, The University of Tokyo, 3-8-1, Komaba, Meguro, Tokyo 153-8902, Japan}



\begin{abstract}

We aim to understand the effect of stellar evolution on the evolution of protoplanetary disks. We focus in particular on the disk evolution around intermediate-mass (IM) stars, which evolve more rapidly than low-mass ones.
We numerically solve the long-term evolution of disks around $0.5$--$5\,\Msun$ stars considering viscous accretion and photoevaporation (PE) driven by stellar far-ultraviolet (FUV), extreme-ultraviolet (EUV), and X-ray emission.
We also take stellar evolution into account and consider the time evolution of the PE rate.
We find that the FUV, EUV, and X-ray luminosities of IM stars evolve by orders of magnitude within a few Myr along with the time evolution of stellar structure, stellar effective temperature, or accretion rate.
Therefore, the PE rate also evolves with time by orders of magnitude, and we conclude that stellar evolution is crucial for the disk evolution around IM stars.

\end{abstract}



\section{Introduction} \label{sec:intro}

So far, the long-term evolution of protoplanetary disks has been mostly investigated by considering viscous accretion and photoevaporation \citep[PE; e.g.,][]{Clarke+01,Alexander+06b,Gorti+09, Owen+10}.
The PE is a thermally driven disk wind from hot disk atmospheres due to the irradiation of high-energy photons \citep[e.g.,][]{Hollenbach+94}, that is, far-ultraviolet (FUV) photons (6--13.6\,eV), extreme-ultraviolet (EUV) photons (13.6--100\,eV) and X-rays ($>100\,$eV).
Most of the previous works, however, do not consider the time evolution of the luminosity of high-energy photons or include all PE mechanisms.

\citet{Gorti+09} investigated the long-term disk evolution considering all PE mechanisms from central stars for the first time.
However, they did not consider the temporal evolution of the luminosity of the EUV and X-rays.
Moreover, the contribution of each mechanism was not clearly shown.
Since the PE rate depends on the UV and X-ray luminosities, it is crucial for disk evolutionary models to adopt realistic models of those luminosities.
We also note that \citet{Alexander+04a} claimed that FUV and EUV from the stellar photosphere are sensitive to the absorption in the stellar atmosphere.

In this paper, we aim to 
(i) investigate the long-term disk evolution (i.e., not the dynamical evolution within several Kepler timescales but the disk evolution for Myr) with realistic FUV, EUV, and X-ray luminosity, considering stellar evolution and the absorption in the stellar atmosphere, 
and (ii) clarify which mechanism of PE plays a dominant role in dispersing disks.

We focus in particular on the influence of stellar evolution.
As we will describe in detail in Sect.\,\ref{sec:starevol}, young stars emit UV photons and X-rays through three mechanisms: magnetic activity, accretion shock, and photospheric radiation. 
Since the magnetic activity originates from the convective motion in the stellar interior, the evolution of the stellar internal structure (i.e., the thickness of the convective envelope, $M\sub{conv}$) is important (see Sect.\,\ref{sec:LXevol}). The $M\sub{conv}$ value of young stars decreases with time, and a radiative core is developed instead.
Moreover, the spectra of photospheric radiation depend on the stellar effective temperature, $\Teff$ (see Sect.\,\ref{sec:spect}).
These quantities, $M\sub{conv}$ and $\Teff$, of young stars evolve on the Kelvin-Helmholtz (KH) timescale, 
\begin{eqnarray}\label{eq:tauKH}
    \tKH
    &=& c\frac{G\Mstar^2}{\Rstar\Lstar} \nonumber\\
    &=& 6.7\,{\rm Myr} \brafracket{\Mstar}{\Msun}^{2} \brafracket{\Rstar}{2\,\Rsun}^{-1} \brafracket{\Lstar}{\Lsun}^{-1} \brafracket{c}{3/7}\,,
\end{eqnarray}
where $\Mstar$ is the stellar mass, $\Rstar$ is the stellar radius and $\Lstar$ is the stellar intrinsic bolometric luminosity, and $c$ is a dimensionless factor that depends on the polytropic index (e.g., $c=3/7$ for fully convective stars and 3/4 for radiative stars).
Given the weak dependence of $\Lstar$ of pre-main-sequence (pre-MS) stars on $\Mstar$ (i.e., roughly $\Lstar\propto\Mstar^2$ for pre-MS stars), 
$\tKH$ of higher-mass stars is shorter; thus, the $\Teff$ and $M\sub{conv}$ of higher-mass stars evolve more rapidly.
Therefore, the PE rate is also expected to change with time, in particular around higher-mass stars.

We note that recent infrared (IR) observations have revealed that the disk evolution around intermediate-mass (IM) stars is different from low-mass stars in the following three respects: the near-IR dust disk lifetime of IM stars is shorter than low-mass stars \citep[][]{Hillenbrand+92,Hernandez+05,Carpenter+06,Yasui+14,Ribas+15}, the H$\alpha$ gas disk lifetime is also shorter \citep{Kennedy+Kenyon09,Yasui+14}, and there is a substantial difference between near- and mid-IR dust disk lifetimes, unlike in low-mass stars \citep[][]{Yasui+14}. Therefore, disk evolution depends on  stellar mass. Following the previous studies above, we define IM as stars of mass above 2--5$\,\Msun$\footnote{We note that \citet{Hernandez+05} and \citet{Ribas+15} defined $>2\,\Msun$ stars as Herbig Ae/Be and high-mass stars, respectively, whereas \citet{Yasui+14} defined 1.5--7$\,\Msun$ stars as IM stars.
}.
There is also a difference in planetary architectures between low-mass and IM stars (i.e., a lack of close-in planets around IM stars), which may result from the different disk evolution \citep[e.g.,][]{Burkert+Ida07,Sato+08,Currie09, Kunitomo+11}. 
To understand these puzzles, as a first step, we investigate the effect of stellar evolution on disk evolution in this paper.

This paper is organized as follows. 
First, we describe our model of the luminosity of the high-energy photons considering stellar evolution.
In Sect.\,\ref{sec:method}, we describe our physical models of the PE and computation method for simulating the disk evolution.
In Sect.\,\ref{sec:results}, we investigate how the disk evolution is affected by stellar evolution.
In Sect.\,\ref{sec:discussion}, we describe the caveats of our model.
The results are summarized in Sect.\,\ref{sec:conclusion}.

\section{Stellar evolution}\label{sec:starevol}

In this section, we first describe the computation methods of the stellar evolution (Sect.\,\ref{sec:starevolcalc}) and stellar atmosphere (Sect.\,\ref{sec:spect}).
Using these results and the observational results, we model the evolution of stellar FUV, EUV, and X-ray luminosity (Sects.\,\ref{sec:LFUV}--\ref{sec:LXevol}).

\subsection{Stellar Evolution Calculation}\label{sec:starevolcalc}

\begin{figure}[!t]
  \begin{center}
    \includegraphics[width=\hsize,keepaspectratio]{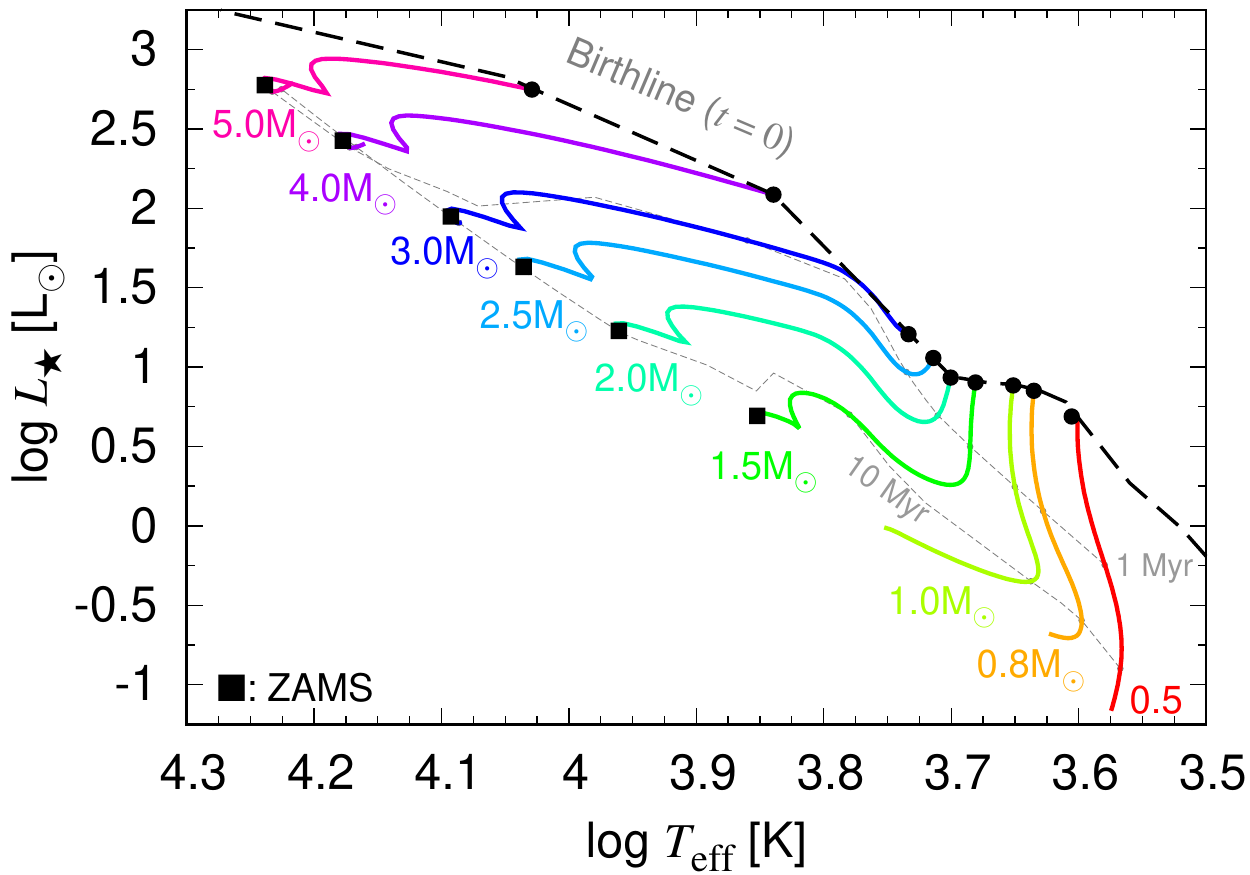}
	\caption{\small{
	Evolutionary tracks of young, 0.5--5\,$\Msun$ stars 
	(from right to left)
	from the birthline \citep[the black dashed line;][]{Stahler+Palla04} to 30\,Myr in the H-R diagram.
	The squares represent the zero-age main sequence.
	The two gray dashed lines show the 1 (top) and 10 (bottom) Myr isochrones.
}}\label{fig:tracks}
    \end{center}
\end{figure}

We simulate the stellar evolution using the code \texttt{MESA} \citep[ver.\,2258, ][]{Paxton+11} \citep[see also][for the details]{Kunitomo+11}.
Figure\,\ref{fig:tracks} shows the evolutionary tracks of $0.5$--$5\,\Msun$ stars in the Hertzsprung-Russell (H-R) diagram.
We assume the solar metallicity.
We adopt the birthline of \citet{Stahler+Palla04} in the H-R diagram as an initial condition.
This corresponds to the standard scenario of star formation (see Sect.\,\ref{sec:caveats-star}).
We note that the luminosity of 0.8--2\,$\Msun$ stars on the birthline is almost the same because of the short thermal timescale, whereas that of $>2\,\Msun$ stars increases with $\Mstar$ because of deuterium burning \citep[][]{Stahler88}.

Here we briefly introduce the basic nature of the stellar pre-MS evolution \citep[see, e.g.,][]{Kippenhahn+Weigert90,Stahler+Palla04}. From the birthline, young low-mass stars evolve along their Hayashi track, which is almost vertical in the H-R diagram due to the strong temperature dependence of the $\rm H^-$ opacity \citep[][]{Hayashi61}. On the Hayashi track, stars are fully convective and shrink due to the radiative energy loss (i.e., the KH contraction).
Since the energy loss results in the increasing internal temperature with time from the virial theorem, and the stellar internal opacity is anticorrelated with temperature (i.e., the Kramers law), the stellar internal opacity decreases with time. Then a radiative core is developed, and a star leaves the Hayashi track. 
We note that high-mass ($>3\,\Msun$) stars are hot enough to have a radiative core from the beginning.
Stars evolve on the horizontal Henyey track, and the stellar effective temperature, $\Teff$, increases with time. The IM pre-MS stars with a high $\Teff$ surrounded by a disk are called Herbig Ae/Be stars \citep[][]{Herbig60,vandenAncker97}.
The evolution of stellar structure and $\Teff$ is a key ingredient in this work (see Sects.\,\ref{sec:spect} and \ref{sec:LXevol}).

For simplicity, we do not consider the $\Mstar$ evolution due to the mass accretion from the disk inner edge to the star or the mass loss via stellar winds \citep[e.g.,][]{Suzuki+13}.

\subsection{Stellar Spectra and Atmospheric Model}\label{sec:spect}

The UV photons are directly emitted from the photosphere of hot IM stars.
Those photons, however, are substantially absorbed in the stellar atmosphere: therefore, the stellar spectra deviate from the blackbody \citep{Alexander+04a}.
Here we quantify the extent of the absorption of UV photons by using a stellar atmospheric model. 
In this subsection, we describe the method and results.

We used version 13.04 of the \texttt{Cloudy} code, last described by \citet{Ferland+13}, to obtain the spectra.
We note that the stellar evolution simulations in \texttt{MESA} do not provide stellar spectra. Therefore, we need to independently calculate the stellar absorption using the \texttt{Cloudy} code.
We adopt the Atlas grids of \citet{Castelli+Kurucz03}, which are available in $\Teff=3500$--50,000\,K, in the case of solar metallicity. We assume $\Rstar = 1\,\Rsun$ and the Stefan-Boltzmann law gives the bolometric luminosity $\Lstar$. We adopt the stellar surface gravity $g=0.33\,\gsun$, where $\gsun=10^{4.44}\,\rm cm/s^2$ is the surface gravity of the Sun.
We note that the results below are not sensitive to the assumed $g$ value (see Appendix\,\ref{sec:Lph-g}) or $\Rstar$.

Figure\,\ref{fig:fred}a shows the stellar spectra as a function of wavelength $\lambda$ in the cases of $\Teff = 20,776, 15,097, 10,128, 7971$ and 3587\,K with and without the absorption in the stellar atmosphere. We note that the latter (i.e., the blackbody spectra) is $\pi\nu B_\nu$, where $\nu$ is the frequency and $B_\nu$ is the Planck function. 
The spectra exhibit  strong absorption at the Lyman break and in the EUV range ($>13.6\,$eV); therefore, we should not use the blackbody for $\PhiEUVph$, as claimed in \citet{Alexander+04a}.
We also find the absorption in the FUV (not only the Ly$\alpha$ absorption at 1216\,\AA) in the case with the low $\Teff$.
We note that we confirmed that the spectrum of a 15,097\,K star is almost the same as Fig.\,1 of \citet{Alexander+04a}.

We define the fraction of the photospheric EUV luminosity $L\sub{EUV,ph}$ to $\Lstar$ as
\begin{equation}
     \fredEUV(\Teff) =  L\sub{EUV,ph}/\Lstar\,.
\end{equation}
We simply assume 50\,eV as the average EUV photon energy and ${\PhiEUVph} = L\sub{EUV,ph}/50\,\rm eV$.
In practice, with the stellar spectra, we calculate $\fredEUV(\Teff)$ by $\fredEUV \equiv \int_{13.6\,\rm\,eV}^{100\,\rm\,eV} F_\nu d\nu /(\SB \Teff^4)$.
We also define $\fredFUV(\Teff) = \LFUVph/\Lstar$, where $\LFUVph$ is the photospheric FUV luminosity.

\begin{figure}[!t]
  \begin{center}
    \includegraphics[width=\hsize,keepaspectratio]{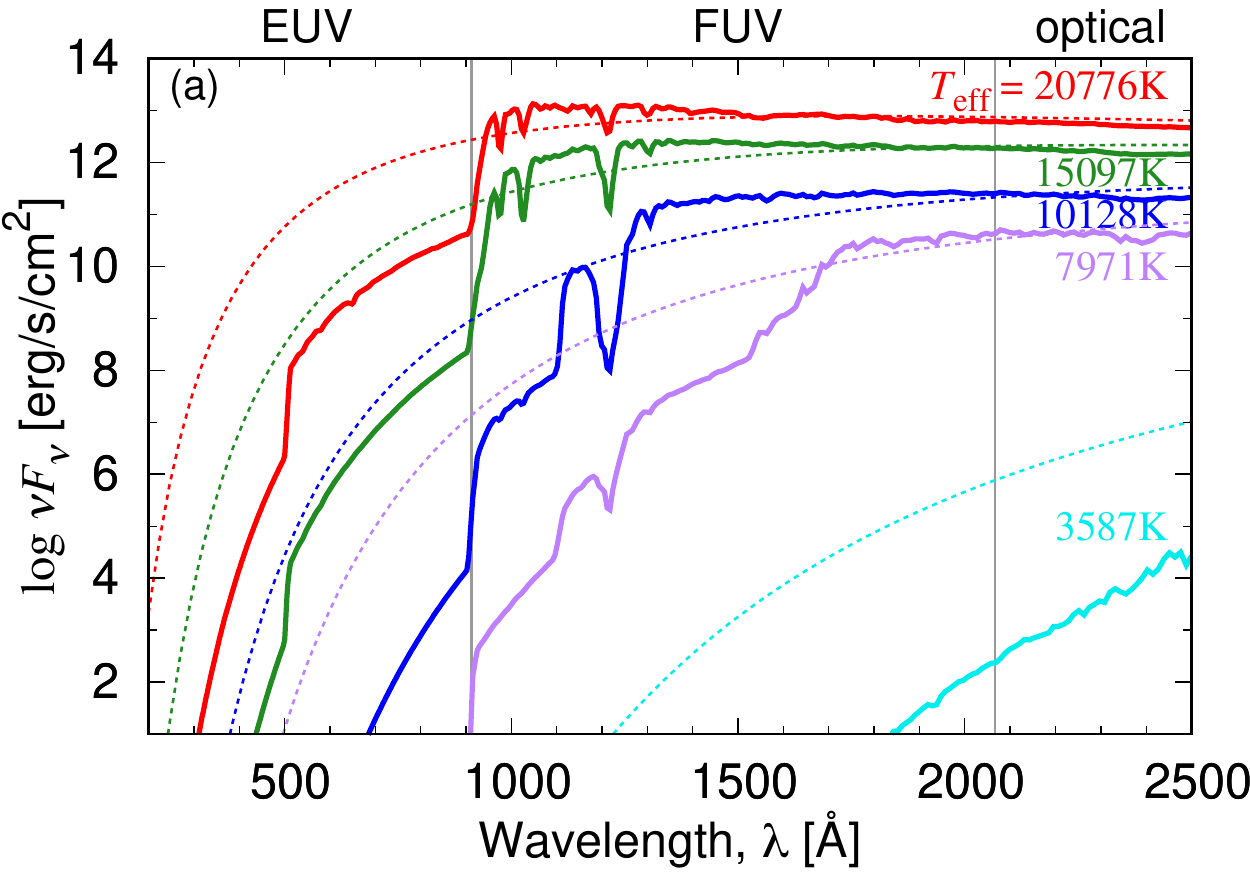}
    \includegraphics[width=\hsize,keepaspectratio]{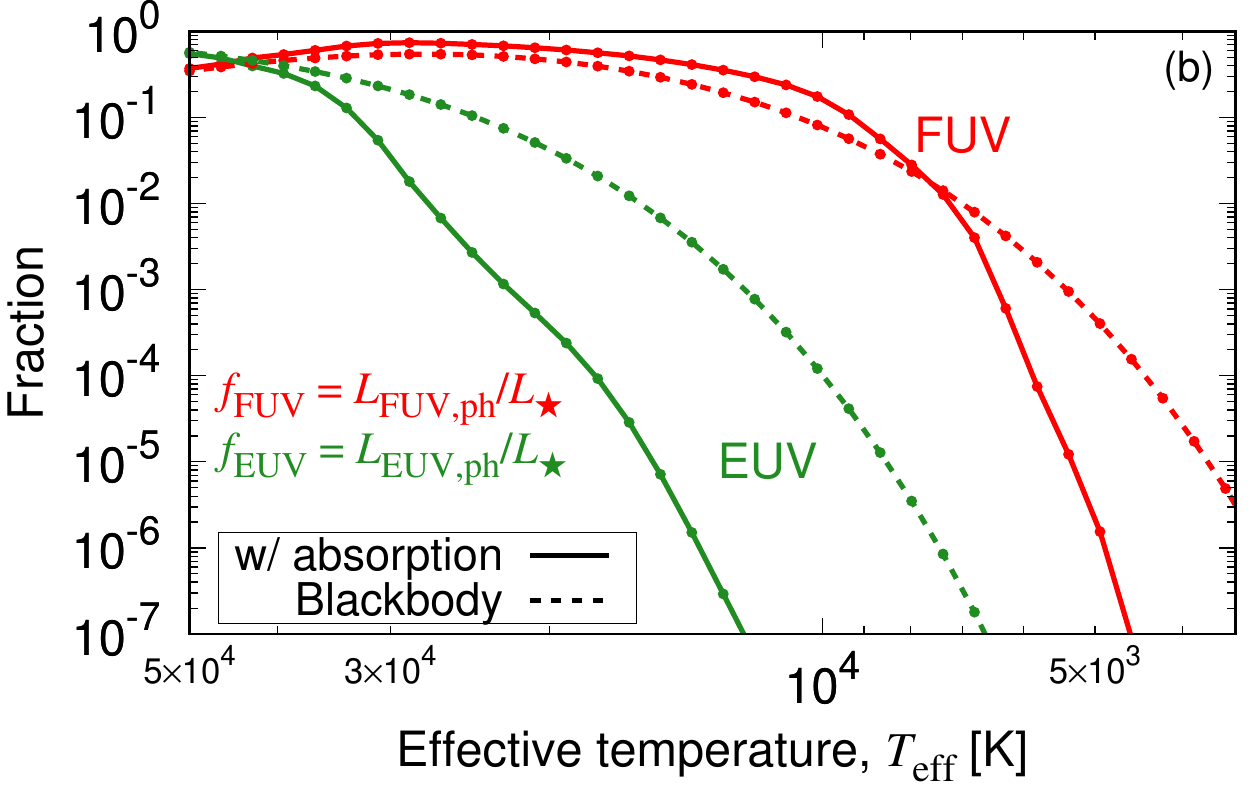}
	\caption{\small{
	The solid and dotted lines show the results with and without the absorption in the stellar atmosphere \citep{Castelli+Kurucz03}, respectively (i.e., the latter is the blackbody spectra), in the case of $g=0.33\,\gsun$.
	(Top panel) Stellar spectra in the cases of $\Teff=20776$, 15097, 10128, 7971 and 3587\,K from top to bottom. The two vertical lines indicate the wavelength at 6 and 13.6\,eV.
	(Bottom) $\fredFUV$ (red) and $\fredEUV$ (green) (see text).
}}\label{fig:fred}
    \end{center}
\end{figure}

Figure\,\ref{fig:fred}b shows the results of $\fredEUV$ and $\fredFUV$ as a function of $\Teff$.
Using the polynomial fitting of Numpy, we obtained the following fitting formulae:
\begin{equation}\label{eq:fredEUV}
    \log \fredEUV = \sum_{i=0}^5 a_i(\log \Teff)^i\,
\end{equation}
in 5000--50,000\,K, where
$a_5 = -95.238145$,
$a_4 =  1998.2116$, 
$a_3 = -16728.880$, 
$a_2 =  69832.410$,
$a_1 = -145282.56$,
$a_0 =  120432.67$,
and 
\begin{equation}\label{eq:fredFUV}
    \log \fredFUV = \sum_{i=0}^6 b_i(\log \Teff)^i
\end{equation}
in the range of $\Teff= 3500$--50000\,K, where 
$b_6 =  177.14306$,
$b_5 = -4452.9922$,
$b_4 =  46546.370$,
$b_3 = -258939.74$,
$b_2 =  808484.59$,
$b_1 = -1343172.0$,
$b_0 =  927492.51$.

We set $\fredEUV=0$ where $\Teff<5\times10^3\,$K and $\fredFUV=0$ where $<3.5\times10^3\,$K.
Together with the evolution of $\Teff$ and $\Lstar$, we obtain the evolution of $\PhiEUVph$ and $\LFUVph$.

\begin{figure}[!t]
  \begin{center}
    \includegraphics[width=\hsize,keepaspectratio]{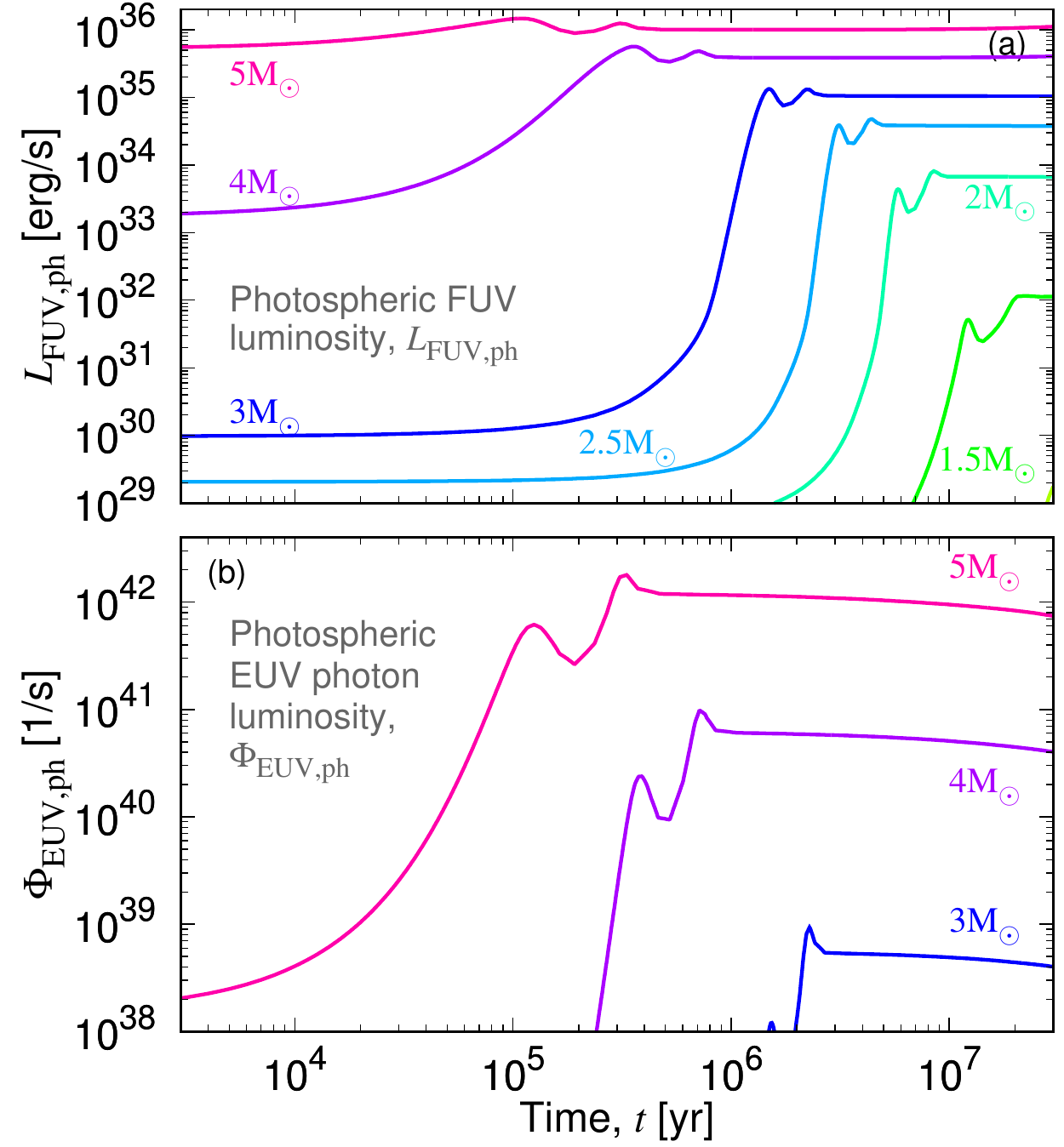}
	\caption{\small{
	Time evolution of the photospheric FUV luminosity, $\LFUVph$, of 1.5--5\,$\Msun$ stars (top panel) and the photospheric EUV photon luminosity, $\PhiEUV$, of 3--5\,$\Msun$ stars (bottom panel).
}}\label{fig:FUVphEUVph}
    \end{center}
\end{figure}

Figure\,\ref{fig:FUVphEUVph} shows the results of the $\LFUVph$ evolution of 1.5--5\,$\Msun$ stars and the $\PhiEUVph$ evolution of 3--5\,$\Msun$.
We have combined $\Teff(t)$ and $\Lstar(t)$ from the stellar evolution simulations (see Sect.\,\ref{sec:starevolcalc}) and the $\fredFUV$ and $\fredEUV$ relations (see the solid lines in Fig.\,\ref{fig:fred}b).
We find that they abruptly increase by orders of magnitude.
Equation\,(\ref{eq:fredFUV}) shows that $\Teff=7342\,\rm K$ is a characteristic temperature; above this temperature, $\fredFUV$ exceeds $10^{-2}$.
We will show the influence of this rapid increase on the disk evolution in Sect.\,\ref{sec:results}.

\citet{Gorti+09} investigated disk evolution including $\LFUVph$ and $\PhiEUVph$.
They adopted the values of MS stars from \citet{Parravano+03}:
$\LFUVph=3.8\times10^{33}, 2.9\times10^{34}, 1.1\times10^{35}, 4.3\times10^{35}$ and $1.3\times10^{36}$\,erg/s for $2, 2.5, 3, 4$ and $5\,\Msun$ stars, whereas $\PhiEUVph=2.4\times10^{42}\,\rm s^{-1}$ for a $5\,\Msun$ star\footnote{
In \citet{Parravano+03}, $\LFUVph$ of  $<1.8\,\Msun$ stars and $\PhiEUVph$ of  $<5\,\Msun$ are not available (see their Table\,1). }
\citep[see also][]{Armitage00}.
These values agree well with the values for MS stars in our model (see Fig.\,\ref{fig:FUVphEUVph}).
We also note that \citet{Parravano+03} indirectly verified their models by comparing them with observed interstellar FUV radiation fields.

\subsection{Stellar FUV Luminosity}\label{sec:LFUV}

Using the results in Sects.\,\ref{sec:starevolcalc} and \ref{sec:spect} and the observational results, we model the stellar FUV luminosity $\LFUV$.
We adopt the same model of $\LFUV$ as \citet{Gorti+09} and assume that $\LFUV$ is the sum of three components,
\begin{equation}\label{eq:LFUV}
    \LFUV = \LFUVacc + \LFUVph + \LFUVcor\,,
\end{equation}
where $\LFUVacc$ originates from the accretion process, and $\LFUVcor$ from the stellar chromosphere.

We assume that 4\% of the gravitational energy of accreting materials ($=G\Mstar\Mdotacc/\Rstar$) is emitted as FUV photons \citep[][]{Calvet+Gullbring98}; therefore,
\begin{equation}\label{eq:LFUVacc}
    \LFUVacc = 10^{-2}\,{\Lsun} 
    \brafracket{\Mstar}{\Msun}
    \brafracket{\Rstar}{\Rsun}^{-1}
    \brafracket{\Mdotacc}{10^{-8}\,\Msun/yr}\,,
\end{equation}
where $\Mdotacc$ is the mass accretion rate onto the star. 
Observations also suggest that the $\LFUV$ of classical T Tauri stars is proportional to $\Mdotacc$ \citep[e.g.,][]{Ingleby+11}.
We also adopt the $\LFUVcor$ model as
\begin{equation}\label{eq:LFUVcor}
    \LFUVcor = 10^{-3.3} \Lstar
\end{equation}
\citep[see section 3.1 of ][and references therein]{Alexander+14}.
We adopt the $\LFUVph$ model in Sect.\,\ref{sec:spect}.
Because $\LFUVacc$ depends on the initial condition and disk evolution, we will show our $\LFUV$ models in Sect.\,\ref{sec:results}.
We note that all the components (i.e., $\LFUVacc$, $\LFUVph$, and $\LFUVcor$) are important (see Fig.\,\ref{fig:t-LFUV}).

\subsection{Stellar EUV Luminosity}\label{sec:PhiEUV}

The origin of EUV photons from pre-MS stars remains unclear, because interstellar hydrogen atoms easily absorb EUV and it is difficult to observationally measure their $\PhiEUV$.
In this paper, we consider EUV from the stellar corona and photosphere and assume that $\PhiEUV$ is the sum of them ($\PhiEUVcor$ and $\PhiEUVph$, respectively) as
\begin{equation}
\Phi\sub{EUV}=\PhiEUVcor+\PhiEUVph\,.   \label{eq:PhiEUV}
\end{equation}
We simply adopt $\Phi\sub{EUV,cor}=10^{41}\,\rm{s}^{-1}$ in this paper (see Sect.\,\ref{sec:caveats}).
We adopt the $\PhiEUVph$ model in Sect.\,\ref{sec:spect}.

\subsection{Stellar X-Ray Luminosity}\label{sec:LXevol}

Stellar X-rays are emitted from the hot corona by magnetic activity. Although the accretion onto the star may also contribute to the X-ray luminosity $\LX$ \citep[see, e.g.,][]{Kastner+02,Kastner+04}, in this paper, we neglect this possibility for simplicity (see Sect.\,\ref{sec:caveats-PE}).
We model the evolution of $\LX$ based on the following two observed features.

First, observations have suggested that $\LX$ depends on the stellar Rossby number.
The $\LX$ of rapid rotators is known to be a function of $\Lstar$; that is, the fractional X-ray luminosity ($R\sub{X}\equiv L\sub{X}/L_{\star}$) is constant at around $10^{-3}$  \citep[e.g.,][]{Vilhu+Rucinski83}.
Most T Tauri stars rotate rapidly and have this relation  \citep[so-called ``saturation'';][]{Flaccomio+03,Preibisch+05,Telleschi+07a}.
On the other hand, $\RX$ of IM stars or slow rotators is more complex. Since the dynamo efficiency depends on both the rotation period and the depth of the convective zone, \citet{Mangeney+Praderie84} and \citet{Noyes+84} introduced the Rossby number, which is the ratio of the rotational period to the convective turnover timescale ($\Ro=P\sub{rot}/\tau\sub{conv}$), as an indicator of the X-ray activity.
\citet{Wright+11} combined the observed data of both saturated and unsaturated stars and derived the following empirical formula: 
$R\sub{X} = \min\left[ 10^{-3.13},5.3\times10^{-6}\,\Ro^{-2.7}\right]$\,.
The threshold value of the saturation is $\Rosat=0.16$.

Second, the $\LX$ of pre-MS IM stars depends strongly on their age.
\citet{Hamaguchi+05} and \citet{Huenemoerder+09} reported that young IM stars on or leaving their Hayashi track have a high $\RX$ ($\sim10^{-3}$--$10^{-4}$)\footnote{
We note that we should be careful with the contribution by an unresolved binary star, but \citet{Hamidouche+08} ruled out this possibility with an 80\% confidence level.}.
On the other hand, the older counterparts, Herbig Ae/Be stars, have smaller values of $R\sub{X}$ ranging from $10^{-5}$ to $10^{-7}$ according to observations \citep[][]{Zinnecker+Preibisch94,Hamaguchi+05,Hamidouche+08,Stelzer+09}.
The strong dependence of the $\LX$ of IM stars on age (or $\Teff$) is shown in \citet{Flaccomio+03}, \citet{Hamaguchi+05}, and \citet{Gregory+16}.
\citet[][see their figure\,9]{Flaccomio+03} showed that the median value of $\LX$ of 2--3\,$\Msun$ stars decreases at around a few Myr by orders of magnitude.
We note that this is consistent with recent observations by \citet{Villebrun+19}, which have suggested that young IM stars possess magnetic fields, whereas most (90\%--95\%) Herbig Ae/Be stars do not.
Therefore, we assume that the evolution of the $\LX$ of IM stars can also be modeled with the Rossby number; the increase of the $\Ro$ number with time results in the $\RX$ decrease.
Although the physical origin of the weak X-ray emission of Herbig Ae/Be stars is still under debate,
we impose a lower limit to $\RX=10^{-7}$ even if $\Ro>4.35\equiv \Rofloor$.

Considering the above observational constraints,
we model the $\LX$ evolution as follows:
\begin{equation}\label{eq:LX}
L\sub{X} = \max \left[ \min\left( 10^{-3.13},5.3\times10^{-6}\,Ro^{-2.7}\right) , 10^{-7}\right]\,L_{\star}\,.
\end{equation}
We note that the choice of the lower limit of $\RX (=10^{-7})$ has little impact on disk evolution.

To compute $\LX$ with Eq.\,\ref{eq:LX}, we need the evolution of $\tau\sub{conv}$ and $\Prot$.
From the mixing-length theory \citep{Cox+Giuli68}, $\tau\sub{conv}$ in the stellar interior can be estimated as
\begin{eqnarray}\label{eq:tauconv}
\tau\sub{conv}&=&\left[ \frac{M\sub{conv}(R_{\star}-R\sub{conv})^{2}}{3L_{\star}} \right]^{1/3}\,,
\end{eqnarray}
where $M\sub{conv}$ is the mass in the convective envelope and $R\sub{conv}$ is the radius at the base of the envelope \citep{Zahn+77,Rasio+96,Villaver+Livio09}.

The $\Prot$ value of stars younger than several Myr (corresponding to the disk lifetime) ranges from 1 to 10\,days and is almost constant with time, probably due to the star-disk locking \citep{Rebull+04,Bouvier08,Gallet+Bouvier13}.
Therefore, we set the fiducial value of $\Prot$ to be 3\,days and investigate the influence of its variation in  Sect.\,\,\ref{sec:Prot}.

\begin{figure}[!t]
  \begin{center}
	\includegraphics[width=\hsize,keepaspectratio]{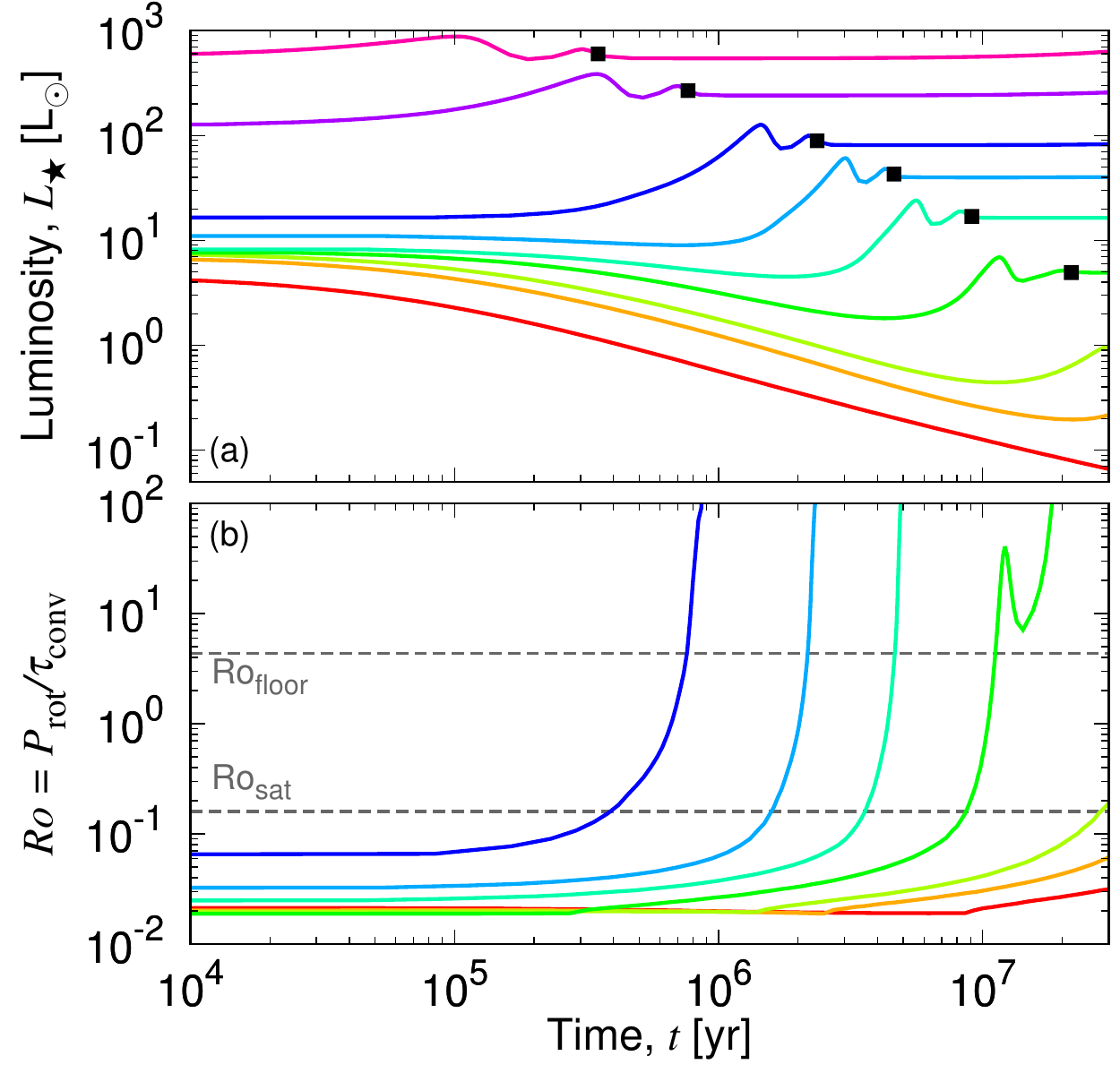}
	\caption{\small{
	Top panel: evolution of $L_\star$ of 0.5--5$\,\Msun$ stars (from bottom to top; color-coded by stellar mass as curves in Fig.\,\ref{fig:tracks}).
	The squares denote the zero-age main sequence.
	Bottom panel: evolution of $\Ro$ ($=P\sub{rot}/\tau\sub{conv}$) of 0.5--3$\,\Msun$ (from bottom to top) with $P\sub{rot}=3\,$days}.
	The dashed lines show the critical Rossby numbers ($\Rosat=0.16$ and $\Rofloor=4.35$).
	We note that 4 and $5\,\Msun$ stars have a large $\Ro$ from the beginning.
}\label{fig:t-L-Mconv}
    \end{center}
\end{figure}

\begin{figure}[!t]
  \begin{center}
    \includegraphics[width=\hsize,keepaspectratio]{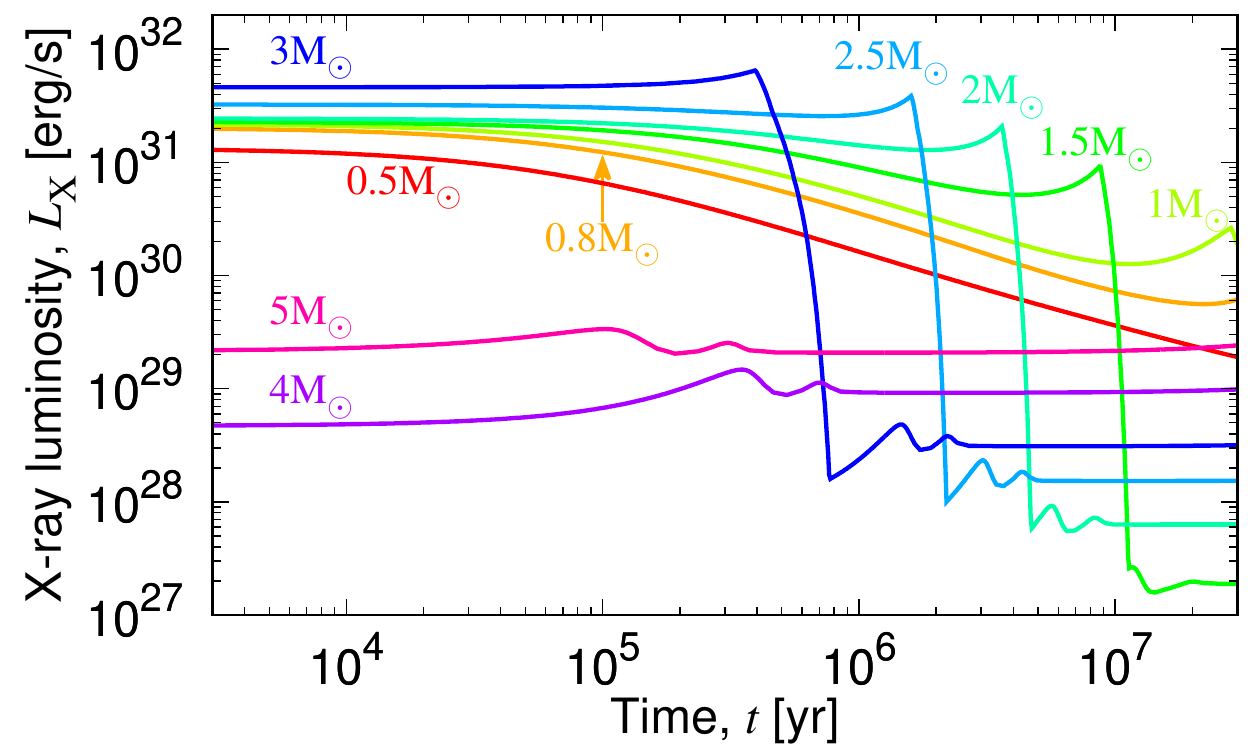}
	\caption{\small{
	Evolution of $\LX$ of 0.5--5\,$\Msun$ stars.
}}\label{fig:t-LX}
    \end{center}
\end{figure}

\begin{figure*}[!t]
  \begin{center}
    \includegraphics[width=\hsize,keepaspectratio]{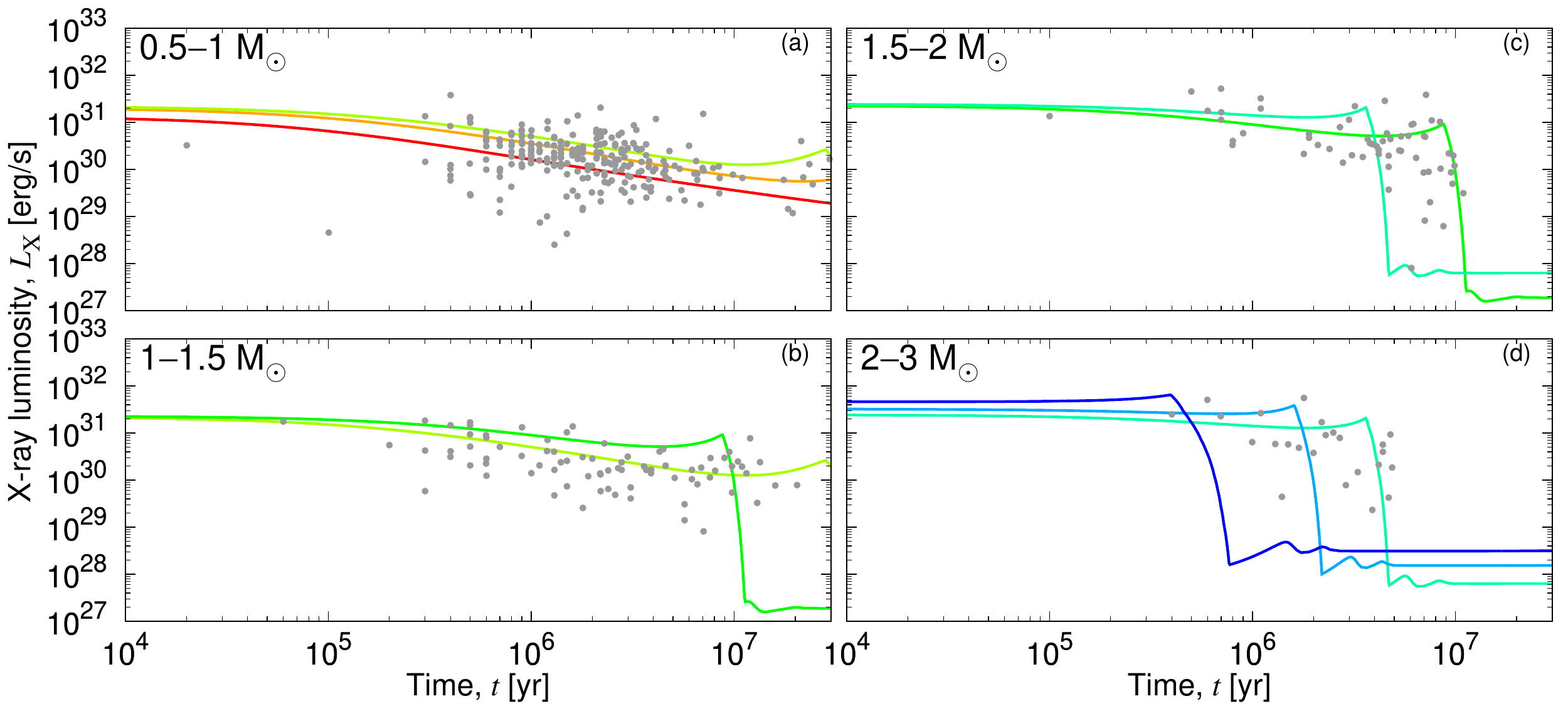}
	\caption{\small{
	Comparison of our $\LX$ evolutionary models of 0.5, 0.8, 1, 1.5, 2, 2.5 and 3$\,\Msun$ stars (solid lines; same as Fig\,\ref{fig:t-LX}) and the observed data \citep[points;][]{Gregory+16} in the range of 0.5--1 (top left), 1--1.5 (bottom left), 1.5--2 (top right) and 2--3 (bottom right) $\Msun$.
}}\label{fig:LX-compobs}
    \end{center}
\end{figure*}

Figures\,\ref{fig:t-L-Mconv}a and \ref{fig:t-L-Mconv}b show the time evolution of $\Lstar$ and $\Ro$, respectively.
Figure\,\ref{fig:t-LX} shows our model of the $\LX$ evolution combining Eq.\,\ref{eq:LX} and Fig.\,\ref{fig:t-L-Mconv}.
One might be skeptical about our prescription of $\LX$. We compare our model of $\LX$ over time with observational data in \citet[][see their figure\,C2]{Gregory+16}.
Figure\,\ref{fig:LX-compobs} shows that our model of the $\LX$ evolution is in good agreement with the data in \citet{Gregory+16}.
The observed $\LX$ data show that 0.5--1$\,\Msun$ stars have a gradual decrease for $\sim10$\,Myr, whereas $\geq1.5\,\Msun$ stars have a decrease by orders of magnitude.
Our model captures such features, and the $\LX$ values and the timing of decrease are also reproduced.
We have also confirmed that our model is consistent with 
\citet{Flaccomio+03}, \citet{Gudel04}, and the Sun\footnote{The Sun has $M\sub{conv}=0.025\,\Msun$ and $R\sub{conv}=0.713\,\Rsun$ \citep{Bahcall+05} and therefore $\tau\sub{conv,\odot}=13.9\,\rm{days}$.
Combining this with $\Prot\simeq26.9\,\rm{days}$, $\Ro_{\odot}\simeq1.94$.
Equation\,(\ref{eq:LX}) with $\Ro_{\odot}=1.94$ gives $\RX=8.9\times10^{-7}$, which is consistent with the observed solar value, $R\sub{X,\odot}\simeq10^{-7}$--$10^{-6}$ \citep{Judge+03}.}.
We admit, however, that Fig.\,\ref{fig:LX-compobs} shows that the $\LX$ values of 1--1.5\,$\Msun$ in our model are several times larger than the median value of the observed $\LX$.
Moreover, the observed $\RX$ has a large scatter \citep[$\sim 1$\,dex; e.g.,][]{Preibisch+05}. We will investigate the impact of the larger/smaller value of $\LX$ in Sect.\,\ref{sec:LX}.

Our stellar evolutionary models described in Sect.\,\ref{sec:starevol} are provided in Table\,\ref{tab:starevol}.

\begin{table*}
 \begin{center}
 \caption{Stellar evolutionary models.  \label{tab:starevol}}
  \begin{tabular}{lllllllllll}
    \tableline\tableline
       $\Mstar$ & $\log t$ & $\Rstar$ & $\Lstar$ & $\Teff$ & $M\sub{conv}$ & $R\sub{conv}$ & $\tau\sub{conv}$ & $\LX$ & $\PhiEUVph$ & $\LFUVph$\\
       $[\Msun]$ & $[\rm{yr}]$ & $[\Rsun]$ & $[\Lsun]$ & $[\rm{K}]$ & $[\Msun]$ & $[\Rsun]$ & $[\rm{days}]$ & $[\rm{erg/s}]$ & $[\rm{1 s^{-1}}]$ & $[\rm{erg/s}]$\\
    \tableline
        \texttt{0.5} & \texttt{0.00} & \texttt{4.537E+00} & \texttt{4.883E+00} & \texttt{4.032E+03} & \texttt{5.000E-01} & \texttt{5.113E-02} & \texttt{1.387E+02} & \texttt{1.391E+31} & \texttt{0.000E+00} & \texttt{1.162E+26} \\
        \texttt{0.5} & \texttt{1.00} & \texttt{4.536E+00} & \texttt{4.686E+00} & \texttt{3.991E+03} & \texttt{4.998E-01} & \texttt{1.858E-01} & \texttt{1.377E+02} & \texttt{1.335E+31} & \texttt{0.000E+00} & \texttt{8.965E+25} \\
        \texttt{0.5} & \texttt{2.00} & \texttt{4.536E+00} & \texttt{4.685E+00} & \texttt{3.991E+03} & \texttt{4.998E-01} & \texttt{1.858E-01} & \texttt{1.377E+02} & \texttt{1.334E+31} & \texttt{0.000E+00} & \texttt{8.964E+25} \\
        \multicolumn{11}{c}{$\vdots$}\\
        \multicolumn{11}{c}{\textit{Continued}}\\
    \tableline
  \end{tabular}
 \end{center}
 \tablecomments{Evolutionary models of young 0.5--$5\,\Msun$ stars. Table\,\ref{tab:starevol} is published in its entirety in the machine-readable format at \href{https://iopscience.iop.org/article/10.3847/1538-4357/abdb2a}{this URL} or \href{https://zenodo.org/record/4595807}{this URL}. A portion is shown here for guidance regarding its form and content.}
\end{table*}

\section{Physical Model and Computation Method of Disk evolution}\label{sec:method}

We simulate the time evolution of protoplanetary disks including the effects of viscous accretion and the time-dependent PE (Sect.\,\ref{sec:basiceq}).
We adopt the PE models from the literature (Sect.\,\ref{sec:PE_model}), considering stellar evolution on the pre-MS (see Sect.\,\ref{sec:starevol}).
The criterion for the disk dispersal is described in Sect.\,\ref{sec:tdisk-condition}.
The numerical method and computational settings are summarized in Sect.\,\ref{sec:num}.

\subsection{Basic Equations}\label{sec:basiceq}
We solve the one-dimensional diffusion equation for the surface density profile \citep[e.g.,][]{Lynden-Bell+Pringle74,Clarke+01}:
\begin{equation} \label{eq:diffusion}
\frac{\partial \Sigma}{\partial t} = \frac{3}{R}\frac{\partial}{\partial R}\left[  \sqrt{R}\frac{\partial}{\partial R}\left( \nu\sub{vis}\Sigma\sqrt{R} \right) \right] - \dot{\Sigma}_{\rm{PE}} (R,t)\,,
\end{equation}
where $\Sigma$ is the surface density, $t$ is the time, $R$ is the distance from the central star, $\nu\sub{vis}$ is the viscosity, and $\dot{\Sigma}_{\rm{PE}}$ is the PE rate, under the cylindrical coordinates ($R, \phi, z$).

We adopt the viscosity model of \citet{Shakura+Sunyaev73}, $\nu\sub{vis} = (2/3) \alpha c_s^2/\Omega$, where $c_s$ is the sound speed at the disk midplane and $\Omega$ is the angular velocity. 
We neglect the disk self-gravity and pressure gradient force and adopt $\Omega = \sqrt{G\Mstar/R^3}$, where $G$ is the gravitational constant.

For the midplane temperature $\Tmid$, we consider both the viscous heating and stellar irradiation, following \citet[][see their Sect.\,2.2]{Kunitomo+20}, which is based on \citet{Suzuki+16}.
Since in this paper, we consider the $\Lstar$ evolution (see Sect.\,\ref{sec:starevol}), the disk temperature in the entire region evolves with time because both viscous heating and stellar irradiation change with time.
We define $\cs^2=\kB\Tmid/(\mu\amu)$, where
$\kB$ is the Boltzmann constant, $\mu=2.34$ is the mean molecular weight, and $\amu$ is the atomic mass unit.

\subsection{PE Models} \label{sec:PE_model}\label{sec:SigdotPE}

In this paper, we consider the PE driven by the irradiation from a central star (so-called ``internal PE'') and do not consider the external irradiation by a nearby massive star \citep[e.g.,][]{Adams+04,Haworth+Clarke19}.

So far, a number of studies have been performed on the internal PE \citep[e.g.,][]{Hollenbach+94,Font+04,Ercolano+08,Gorti+Hollenbach09, Tanaka+13, Komaki+20}.
We also refer the reader to recent reviews such as \citet{Alexander+14}, \citet{Gorti+16}, and \citet{Ercolano+Pascucci17}.
We consider the PE driven by FUV, EUV, and X-rays and we adopt their mass-loss rates from the literature.
We assume that the dominant heating source among the three at the wind launching region determines the mass-loss rate $\SigdotPE$, and therefore 
\begin{equation}\label{eq:SigdotPE}
    \SigdotPE(R,t) = \max \braa\SigdotFUV(R,t), \SigdotEUV(R,t), \SigdotX(R,t)\kett\,,
\end{equation}
where $\SigdotFUV, \SigdotEUV$ and $\SigdotX$ are the PE rate driven by FUV, EUV, and X-rays, respectively.
We note that one might think that $\SigdotPE$ can be proportional to the total energy deposited in the disk atmosphere, and therefore $\SigdotPE = \SigdotFUV + \SigdotEUV + \SigdotX$. We have confirmed that the two expressions of $\SigdotPE$ make little difference in the results ($<8\%$ in disk lifetime) because one process among the three almost always dominates.

The PE rate has two regimes: 
one is for primordial disks, and the other is for disks with an inner hole.
In the latter, the outer disk is directly irradiated, and therefore the PE profile is changed (so-called ``direct PE'').
We consider both regimes.

\begin{figure}[!t]
  \begin{center}
    \includegraphics[width=\hsize,keepaspectratio]{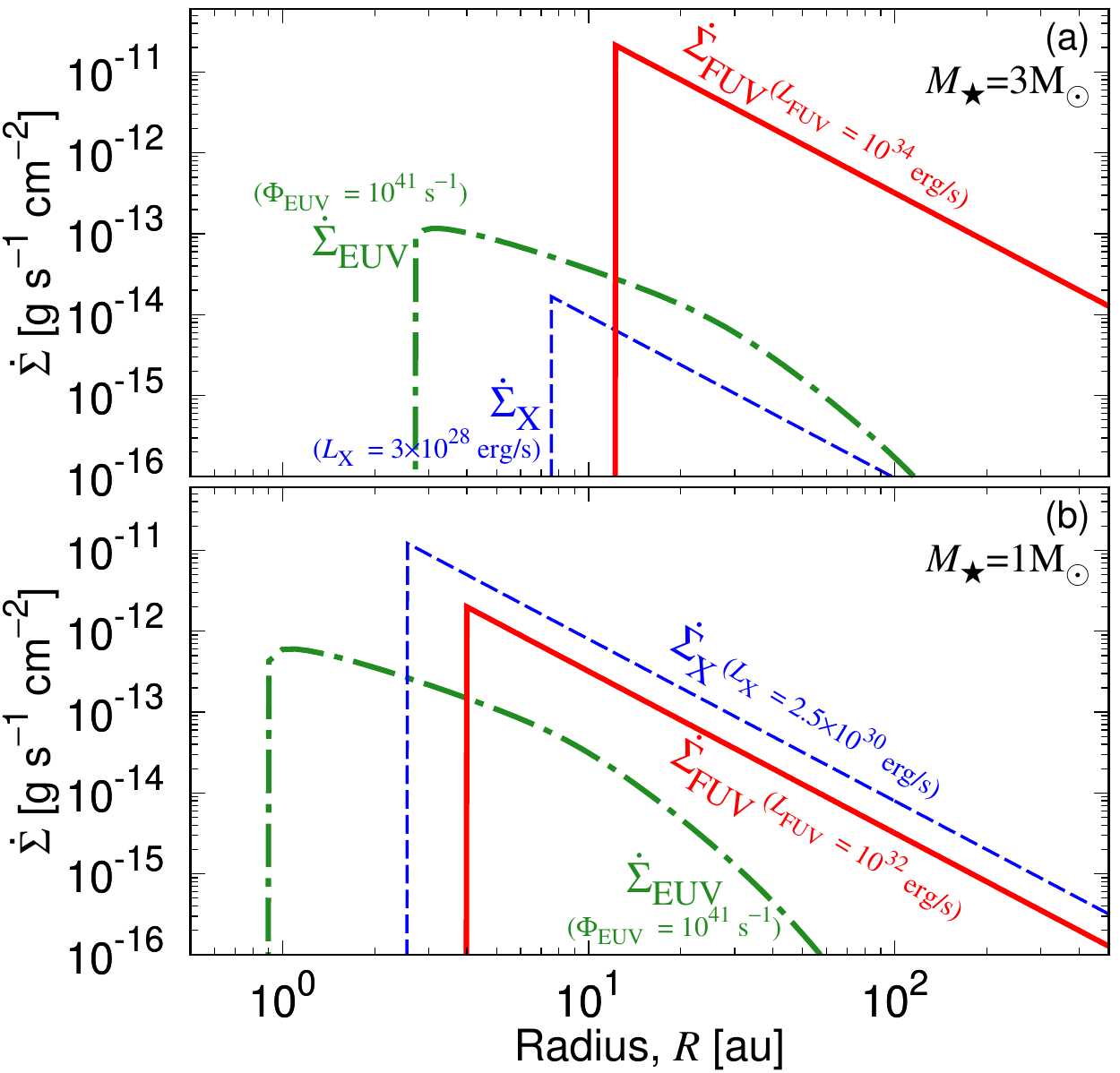}
	\caption{\small{
	Examples of mass-loss profiles by FUV (solid red; $\SigdotFUV$), EUV (dotted-dashed green; $\SigdotEUV$) and the X-ray (dashed blue; $\SigdotX$) in the case of a disk without an inner hole.
	The top and bottom panels show the cases
	around a $3\,\Msun$ star with $\LFUV=10^{34}$\,erg/s, $\PhiEUV=10^{41}\,\rm{s^{-1}}$, and $\LX=3\times10^{28}$\,erg/s,
	and around a $1\,\Msun$ star with $\LFUV=10^{32}$\,erg/s, $\PhiEUV=10^{41}\,\rm{s^{-1}}$, and
	$\LX=2.5\times10^{30}$\,erg/s, respectively.
	Since those luminosities evolve with time, these profiles are just an example.
}}\label{fig:SigdotPE}
    \end{center}
\end{figure}

We adopt the same $\SigdotEUV$ model as in \citet{Kunitomo+20}; 
$\SigdotEUV$ for primordial disks in \citet{Alexander+Armitage07} and that for the direct PE in \citet{Alexander+06a}.
The total mass-loss rates for the EUV PE in both regimes are
\begin{equation}
\Mdot\sub{EUV,p} = 1.6\times 10^{-10}\,{\Msun/{\rm{yr}}}
\brafracket{\Phi\sub{EUV}}{10^{41}\,{\rm{s}}^{-1}}^{1/2}
\brafracket{\Mstar}{1\,\Msun}^{1/2} \label{eq:EUVprim}
\end{equation}
and
\begin{equation}
\Mdot\sub{EUV,d} = 1.3\times 10^{-9}\,{\Msun/{\rm{yr}}}
\brafracket{\Phi\sub{EUV}}{10^{41}\,{\rm{s}}^{-1}}^{1/2} \brafracket{R\sub{hole,EUV}}{3\,\au }^{1/2}\,, \label{eq:EUVdirect}
\end{equation}
where $\Phi\sub{EUV}$ is the EUV photon luminosity and $R\sub{hole,EUV}$ is the inner hole size for the EUV. 
We assume the aspect ratio $h/R=0.05$ in Eq.\,(\ref{eq:EUVdirect}) \citep[see][]{Alexander+06a}, where $h=\sqrt{2}c_s/\Omega$ is the gas scale height\footnote{
We note that the factor of $\sqrt{2}$ is sometimes not included. We include it following \citet{Kunitomo+20}.
}.
The $\SigdotEUV$ profile of primordial disks has a peak at $\simeq1\,{\rm{au}}\,(M_{\star}/\Msun)$.
We refer the reader to \citet{Alexander+Armitage07} for the full formula of $\SigdotEUV$ (see also Fig.\,\ref{fig:SigdotPE}).

As for the X-ray PE rate, the prescription in \citet[][]{Owen+12} has been widely used.
In the case of a $1\,\Msun$ star with $\LX=10^{30}\,\rm  erg/s$, $\SigdotX$ has a peak value 
$(\equiv \SigdotoX = 5.1\times10^{-12}\,{\rm g\,s^{-1}\,cm^{-2}})$ at $2.5\,{\rm au}\,(\Mstar/\Msun)$\footnote{
The peak of the X-ray PE (at 2.5\,au for a $1\,\Msun$ star) is farther than that of the EUV PE (at 1\,au) because the X-ray PE is launched from the atomic layer ($\simeq3000$--5000\,K), whereas the EUV PE is from the $10^4$\,K layer \citep{Alexander+14}.
}, decreases with radius as $R^{-2}$, and has a sharp cutoff at several tens of au.
The cutoff is, however, not seen in the recent study by \citet[][see their Fig.\,5]{Picogna+19}.
For the primordial disks, we adopt 
\begin{equation}\label{eq:SigdotX}
    \SigdotX = \SigdotoX
    \brafracket{\LX}{10^{30}\,{\rm erg/s}}
    \brafracket{R}{ 2.5\,{\rm au} }^{-2}\,,
    \label{eq:XPE}
\end{equation}
outside $2.5\,{\rm au}(\Mstar/\Msun)$. In the inner region, the disk gas is gravitationally bound to the disk and does not flow out (i.e., $\SigdotX=0$).
We note that we neglect the weak dependence on stellar mass ($\propto\Mstar^{-0.068}$) in the original $\SigdotX$ profile in \citet{Owen+12}.
We note that the X-ray PE rate in \citet{Owen+12} has recently been called into question; the radiation-hydrodynamic (RHD) simulations with self-consistent thermochemistry by \citet{Wang+Goodman17} and \citet{Nakatani+18b} disagree with the results in \citet{Owen+12}, and therefore Eq.\,(\ref{eq:XPE}) may overestimate the X-ray PE rate \citep[see also a pioneering study by ][]{Gorti+Hollenbach09}.
We will discuss this issue in Sect.\,\ref{sec:caveats-PE}.

For $\SigdotX$ of the direct PE, we adopt the model in \citet[][see their appendix B]{Owen+12}, which peaks at the inner edge of the outer disk. The total mass-loss rate is
\begin{equation}
\Mdot\sub{X,d} = 4.8\times 10^{-9}\,{\Msun/\rm{yr}}
\brafracket{\LX}{10^{30}\,\rm{erg/s}}^{1.14}
\brafracket{\Mstar}{\Msun}^{-0.148}\,, \label{eq:Xdirect}
\end{equation}
where the subscript ``d'' stands for the direct PE.

We need to define the inner hole sizes for the direct PE for the EUV and X-rays. We also modify $\SigdotEUV$ and $\SigdotX$ for the direct PE to avoid numerical problems using ``smoothing functions.''
We refer readers to \citet[][see their Sect.\,2.4]{Kunitomo+20} for the full details of these prescriptions.

As for the $\SigdotFUV$ model, no formula is available to date in the literature.
We construct the $\SigdotFUV$ model as a function of $R$ and the stellar FUV luminosity $\LFUV$, based on the results in \citet{Gorti+Hollenbach09} and \citet{Wang+Goodman17}.
The latter performed RHD simulations, whereas the former conducted calculations using a hydrostatic model.

\citet{Gorti+Hollenbach09} investigated the dependence of $\SigdotFUV$ on $\LFUV$ (see model F10, S and F0.1 in their figure\,4) around a $1\,\Msun$ star.
The $\SigdotFUV$ value beyond $4\,\au$ changes by about 1 order of magnitude by varying $\LFUV$ by an order of magnitude.
Therefore, we assume that $\SigdotFUV\propto\LFUV$ and the FUV PE mass loss occurs beyond $4\,\au\,(\Mstar/\Msun)$.
We note that $4\,\au$ corresponds to the critical radius for $\simeq2000\,\K$ gas around a $1\,\Msun$ star \citep[][]{Liffman03}. 
The gas heated by FUV is much cooler than that by EUV, which is $\simeq10^4\,\K$ \citep[see also][]{Nakatani+18a}.

The $\SigdotFUV$ profile in \citet{Gorti+Hollenbach09} is a complex function of $R$ (see their figure\,4), whereas \citet{Wang+Goodman17} claimed that $2\pi R^2\SigdotFUV$ is almost constant (see their figure\,8).
\citet[][see their Sect.\,5.2]{Wang+Goodman17} confirmed that the difference arises from how to estimate the mass-loss rate; the sonic point is different between the hydrodynamic simulations in \citet{Wang+Goodman17} and the hydrostatic models in \citet{Gorti+Hollenbach09}.
The $\SigdotFUV\propto R^{-2}$ profile seems energetically reasonable.
Therefore, from the results in \citet{Wang+Goodman17}, we assume that $\SigdotFUV\propto R^{-2}$ and $\SigdotFUV = 10^{-12}\,\rm g\,cm^{-2}\,s^{-1} (\equiv \Sigdoto)$ at 4\,au around a $1\,\Msun$ star.
As a result, we adopt the following $\SigdotFUV$ profile: in the outer disk beyond $4\,\au\,(\Mstar/\Msun)$,
\begin{equation}\label{eq:SigdotFUV}
    \SigdotFUV = \Sigdoto
    \brafracket{\LFUV}{10^{31.7}\,{\rm erg/s}}
    \brafracket{R}{ 4\,{\rm au} }^{-2}\,,
    \label{eq:FUVPE}
\end{equation}
and in the inner disk ($R<4\au\,(\Mstar/\Msun)$), $\SigdotFUV = 0$.

Figure\,\ref{fig:SigdotPE} shows examples of the $\SigdotPE$ profiles of two cases; one is 
around a $3\,\Msun$ star with 
$\LFUV=10^{34}$\,erg/s, 
$\PhiEUV=10^{41}\,\rm{s^{-1}}$, and $\LX=3\times10^{28}$\,erg/s,
and the other is 
around a $1\,\Msun$ star with 
$\LFUV=10^{32}$\,erg/s, 
$\PhiEUV=10^{41}\,\rm{s^{-1}}$, and
$\LX=10^{31}$\,erg/s.
As described in Sect.\,\ref{sec:starevol}, these luminosities evolve with time, and therefore the PE rate varies with time.

\subsection{Disk dispersal criterion}\label{sec:tdisk-condition}

In this study, we define the time when the disk mass, $\Md$, decreases down to $10^{-8}\,\Mdini$ as the disk lifetime, $\tdisk$, where $\Mdini$ is the initial disk mass.
Here we take a numerical factor $10^{-8}$ but $\tdisk$ is insensitive to it, if it is $\leq10^{-4}$.

We note that \citet{Kimura+16} and \citet{Kunitomo+20} measured the inner disk lifetime when the optical depth of the inner disk (i.e., the IR-emitting region) becomes unity.
Considering the fact that the IR is emitted by dust grains that are not modeled in this study (see Sect.\,\ref{sec:caveats-dust}), here we measure $\tdisk$ using $\Md$. However, we note that the inner disk lifetime using the optical depth is almost the same as $\tdisk$ in this study, because an entire disk disperses quickly once a gap opens (see Fig.\,\ref{fig:3Msun}a).

\subsection{Numerical method}\label{sec:num}

We numerically solve Eq.\,(\ref{eq:diffusion}) using the time-explicit method based on \citet{Kunitomo+20}.
The calculation domain ranges from 0.01 to $10^4\,\au$. The grid size is in proportion to $\sqrt{R}$ and the number of mesh points is 2000. 
The zero-torque boundary condition is imposed at both the inner and outer boundaries.
We measure $\Mdotacc$ at the innermost cell.
We stop calculations when the disk is completely dispersed.

We adopt the self-similar solution \citep{Lynden-Bell+Pringle74} as an initial surface density profile given by
\begin{equation}\label{eq:Sigini}
\Sigma (R,t=0) = \frac{\Mdini}{2\pi R_1^2}\frac{\exp{(-R/R_1)}}{R/R_1}\,.
\end{equation}
The characteristic radius $R_1$ represents the location outside which the $e^{-1}$ of the disk mass resides.

We choose input parameters to reproduce observational constraints as summarized in Table\,\ref{tab:input}. First, from the observed relation that disk masses are proportional to $\Mstar$ \citep[e.g.,][]{Williams+Cieza11,Andrews+13,Mohanty+13,Pascucci+16}, 
we adopt
\begin{equation}\label{eq:Mdini}
\Mdini\propto\Mstar\,.
\end{equation}
The proportionality factor ranges from 0.001 to 0.1. Given that this value decreases with time, we start calculations with a massive disk, $M_{\rm{d,ini}}=0.1\,\Mstar$ (i.e., from the early phase).
We note that the quantity of $\Mdini/\Mstar$ does not change the qualitative results on the disk lifetimes.

Second, following \citet{Gorti+09}, we adopt
\begin{equation}\label{eq:alpha}
\alpha\propto \Mstar
\end{equation}
in order to reproduce the observed relation $\Mdotacc \propto\Mstar^{2}$ \citep[e.g.,][]{Calvet+04,Muzerolle+05}.
We assume that magnetorotational instability \citep[MRI;][]{Velikhov59,Chandrasekhar61,Balbus+Hawley91} is the source of the turbulent viscosity, and we adopt $\alpha=10^{-2}\,(\Mstar/\Msun)$.
Equation\,(\ref{eq:alpha}) is derived with the following assumptions: the steady-state accretion ($\dot{M}\sub{acc}=3\pi\Sigma\nu\sub{vis}$), the constant $R_1$ with $\Mstar$ and Eq.\,(\ref{eq:Mdini}) (therefore $\Sigma\propto\Mstar$), optically-thin disk temperature \citep[see, e.g., Eq.\,6 of][]{Kunitomo+20}, $L_{\star}\propto\Mstar^2$ \citep[see $\Lstar$ at 1\,Myr in Fig.\,\ref{fig:t-L-Mconv}a or][]{Siess+00}, and Keplerian $\Omega$.

Finally, we adopt the initial disk radius $R_{1}=50\,\au$.
\citet{Andrews+10} measured dust disk radii from millimeter-wavelength observations and found that they range from 14 to 200\,au and peak at $\sim30\,$au (see their figure 3).
Considering that recent studies have suggested that gas disks are likely to be larger than dust disks \citep[e.g.,][]{Ansdell+18}, we adopt $R_{1}=50\,\au$ in this paper.
\citet{Andrews+10} did not find a clear correlation between the disk radius and $\Mstar$ \citep[see also][]{Ansdell+18,Long+19}. Although \citet{Andrews+18} recently suggested a weak correlation with $\Mstar$, in this paper, we adopt the constant $R_1$ with $\Mstar$ for simplicity.

\begin{table}
 \begin{center}
 \caption{Fiducial Disk Model.  \label{tab:input}}
  \begin{tabular}{lc}
    \tableline\tableline
       Parameter & Value \\
    \tableline
    Initial disk mass $M\sub{d,ini}$ & $0.1\,M_\star$\\
    Viscosity parameter $\alpha$ & $10^{-2}(M_\star/\Msun)$\\
    Initial characteristic radius $R_{1}$& 50\,\au\\
    Coronal EUV luminosity $\Phi_{\rm{EUV,cor}}$ & $10^{41}\,{\rm{s^{-1}}}$\\
    \tableline
  \end{tabular}
 \end{center}
\end{table}

\section{RESULTS}\label{sec:results}

\subsection{Overview of Disk Evolution}
\label{sec:overview}

\begin{figure*}[!t]
  \begin{center}
    \includegraphics[width=\hsize,keepaspectratio]{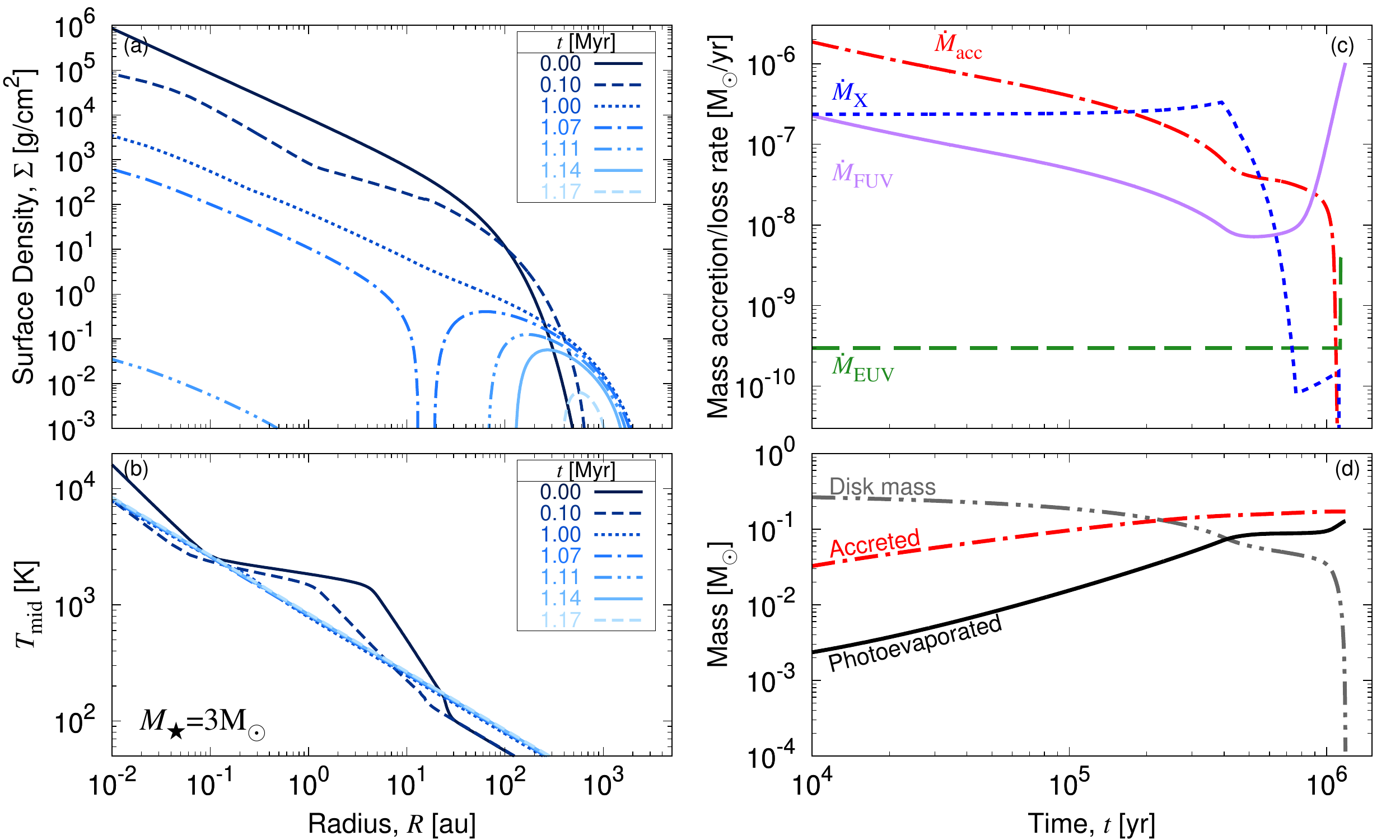}
    \caption{\small{
    Temporal evolution of a disk around a $3\,\Msun$ star. 
	Panels (a) and (b) show the evolution of the surface density ($\Sigma$) profile and the midplane temperature ($\Tmid$) profile, respectively. Each line shows a snapshot between zero and 1.17\,Myr. 
	Panel (c) shows the evolution of the mass accretion rate ($\Mdotacc$; dotted-dashed line) and the mass-loss rate by the X-ray ($\MdotX$; dotted line), EUV PE ($\MdotEUV$; dashed line) and FUV ($\MdotFUV$; solid line) PE. 
	Panel (d) shows the evolution of disk mass ($\Md$; double dotted-dashed line), the time-integrated masses of accretion ($\Macc$; dotted-dashed line) and photoevaporation ($\MPE$; solid line).
}}\label{fig:3Msun}
    \end{center}
\end{figure*}

In this subsection, we show the disk evolution around a $3\,\Msun$ star with the fiducial settings listed in Table\,\ref{tab:input}. 
In our results, $t=0$ corresponds to the time when stars appear on their birthline.
We consider the three PE mechanisms: FUV, EUV, and X-rays.
In the four panels of Fig.\,\ref{fig:3Msun}, we show the evolution of (a) the $\Sigma$ profile, (b) the $\Tmid$ profile, (c) $\Mdotacc$ and the mass-loss rates, 
and (d) the time-integrated accreted or lost mass.
We define $\MdotFUV\equiv\int 2\pi R\SigdotFUV dR$ and $\MdotX\equiv\int 2\pi R\SigdotX dR$ (see Eqs.\,\ref{eq:SigdotX} and \ref{eq:SigdotFUV}).
Both are integrated over the entire computation domain.
The time-integrated accreted mass is $\Macc\equiv \int \Mdotacc dt$, and the total mass lost by the PE is $\MPE\equiv\int \SigdotPE dRdt$ (see Eq.\,\ref{eq:SigdotPE}).
We note that we have checked the mass conservation in our simulations:  $\Mdini=\Md(t)+\Macc(t)+\MPE(t)$ with a precision of $<10^{-10}$.

The qualitative behavior of the evolution in Fig.\,\ref{fig:3Msun} is the same as the results in previous works \citep[e.g.,][]{Clarke+01,Alexander+06b,Gorti+09,Owen+10}:
(i) the $\Md$ decreases with time due to viscous accretion, (ii) a gap is created when and where the accretion rate decreases down to the PE rate, (iii) the inner disk depletes in the viscous timescale at the gap, and (iv) after the dispersal of the inner disk, the outer disk is directly irradiated and also quickly dispersed.
The gap opens at $\sim15$\,au, slightly outside the peak of $\SigdotFUV$ (see Sect.\,\ref{sec:PE_model}). 
We note that the period of the phase (iii) is consistent with the viscous timescale, $\tvis$, at the gap given by
\begin{eqnarray}
    &&\tvis \equiv \frac{R^2}{\nu\sub{vis}} \\
    &&= 0.07\,\mathrm{Myr}
    \brafracket{R}{15\,\mathrm{au}}^{1/2}
    \brafracket{\Tmid}{100\,\mathrm{K}}^{-1}
    \brafracket{\Mstar}{3\,\Msun}^{1/2}
    \brafracket{\alpha}{0.03}^{-1}\,.\nonumber
\end{eqnarray}
We note that the nonsmooth $\Tmid$ profile in Fig.\,\ref{fig:3Msun}b results from the nonlinear function of the opacity \citep[see][]{Kunitomo+20}.

\begin{figure}[!t]
  \begin{center}
    \includegraphics[width=\hsize,keepaspectratio]{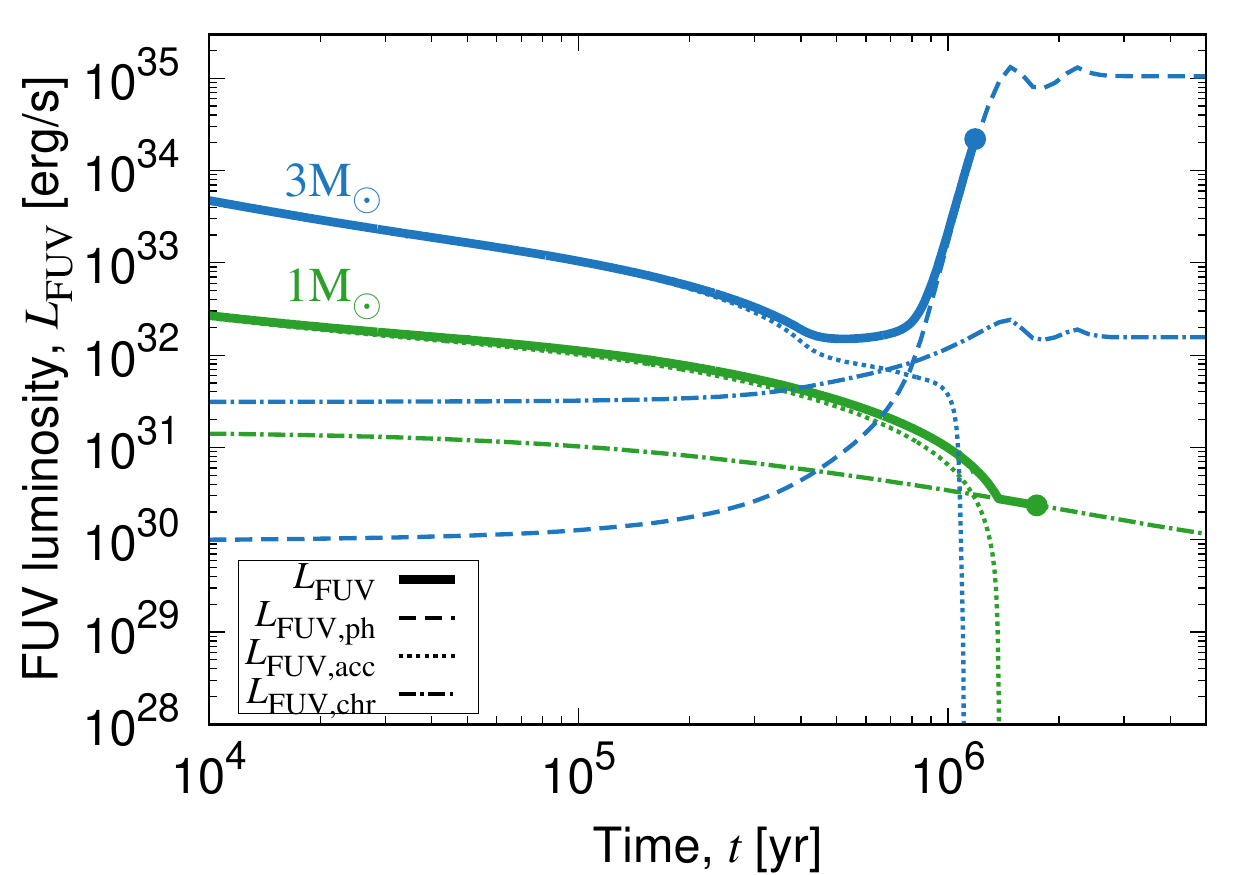}
	\caption{\small{
	Temporal evolution of $\LFUV$ (solid lines), $\LFUVph$ (dashed lines), $\LFUVacc$ (dotted lines) and $\LFUVcor$ (dotted-dashed lines) in the cases of $\Mstar=3$ (blue) and 1 (green) $\Msun$.
	The disks disperse and the simulations stop at the filled circles.
	We note that the $\LFUVph$ of a $1\,\Msun$ star is negligibly low.
}}\label{fig:t-LFUV}
    \end{center}
\end{figure}

Figure\,\ref{fig:3Msun}c shows that the mass-loss rates evolve with time, unlike the previous studies.
Although the X-ray PE rate, $\MdotX$, is high (a few $10^{-7}\,\Msun/\yr$) in the early phase, $\MdotX$ decreases by more than 3 orders of magnitude between 0.4 and 0.8\,Myr.
This is induced by stellar evolution; at this phase, a $3\,\Msun$ star develops a large radiative core, $\Ro$ increases, and therefore $\LX$ and $\MdotX$ decrease (see Figs.\,\ref{fig:t-L-Mconv}b and \,\ref{fig:t-LX}).
Instead, $\MdotFUV$ rapidly increases by more than 1 order of magnitude at $\simeq1$\,Myr.
This is because, after $\simeq1$\,Myr, $\Teff > 7300\,$K and $\LFUVph/\Lstar> 10^{-2}$; that is, the stellar surface becomes hot enough to emit FUV from the photosphere.
We stress that although \citet{Gorti+09} already found that the rapid disk dispersal around IM stars is induced by the PE driven by photospheric UV, they did not consider stellar evolution (see Sect.\,\ref{sec:importance}).
Since the rapid increase of $\MdotFUV$ has a strong impact on the disk evolution, we claim that stellar evolution is crucial for the disk dispersal around IM stars.

Figure\,\ref{fig:t-LFUV} shows the evolution of $\LFUV$ in the cases of $\Mstar=1$ and $3\,\Msun$.
The disk evolution around a $1\,\Msun$ star is shown in Appendix\,\ref{app:lowmass}.
In the $1\,\Msun$ case, $\LFUVacc$ dominates in almost the entire phase, which is consistent with observations \citep[see, e.g., ][]{Ingleby+11} and previous theoretical study \citep{Gorti+09}.
Along with the decrease in $\Mdotacc$, $\LFUV$ decreases with time, and in the late phase, $\LFUVcor$ dominates.
In the case of $3\,\Msun$ stars, however, although $\LFUVacc$ dominates in the early phase, $\LFUVph$ rapidly increases by orders of magnitude as $\Teff$ increases at $\simeq1$--1.5\,Myr.
We note that, in the $4\,\Msun$ case, the switch occurs at $4\times10^4\,$yr, and in the $5\,\Msun$ case, $\LFUVph$ always dominates.

We note that the initial value of $\LFUVacc$ of the $3\,\Msun$ star is $\simeq1$ order of magnitude larger than that of $1\,\Msun$.
This is because we adopt the initial condition to reproduce the observed relation $\Mdotacc\propto\Mstar^2$ (see Sect.\,\ref{sec:num}).

\subsection{Importance of Stellar Evolution}\label{sec:importance}

\begin{figure}[!t]
  \begin{center}
    \includegraphics[width=\hsize,keepaspectratio]{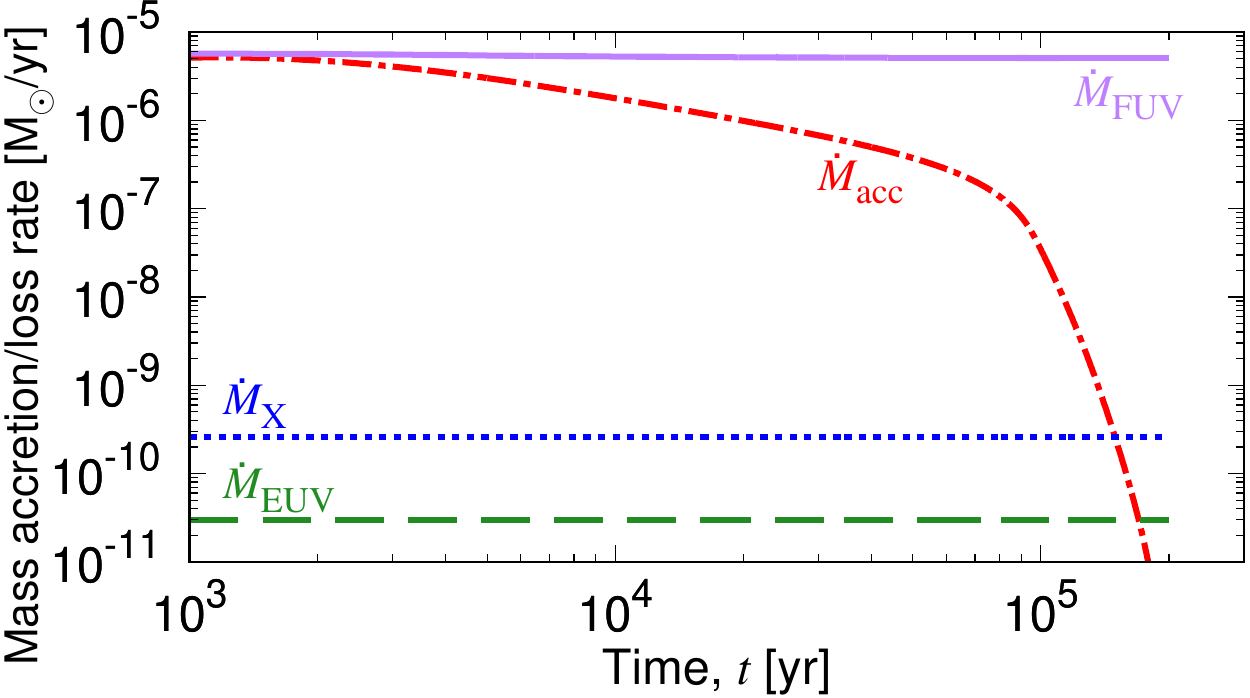}
	\caption{\small{ 
	Same as Fig.\,\ref{fig:3Msun}c but with the constant $\LFUVph (=1.1\times10^{35}\,{\rm erg/s}), \PhiEUV (=1.0\times10^{39}\,{\rm erg/s})$, and $\LX (=5.0\times10^{28}\,{\rm erg/s})$ with time.
	Although $\MdotFUV$ changes with time due to the $\LFUVacc$ evolution, it is negligibly low.
}}\label{fig:importance}
    \end{center}
\end{figure}

In Sect.\,\ref{sec:overview}, we showed that the photospheric FUV radiation has a dominant role in the disk dispersal around a $3\,\Msun$ star.
We again note that it had already been found by \citet{Gorti+09}, and as an update from their study, we considered the stellar evolution.
To illustrate its importance, we performed the same simulation of Fig.\,\ref{fig:3Msun} but without the time evolution of $\LFUVph, \PhiEUV$ and $\LX$ as in \citet{Gorti+09}.
We adopt 
$\LFUVph=1.1\times10^{35}\,{\rm erg/s}$, 
$\PhiEUV =1.0\times10^{39}\,{\rm erg/s}$, and 
$\LX =5.0\times10^{28}\,{\rm erg/s}$ following \citet{Gorti+09} (see also Sect.\,\ref{sec:starevolcalc}).
Figure\,\ref{fig:importance} shows that $\MdotFUV$ is kept high from the beginning; therefore, the disk disperses much earlier than the case in Fig.\,\ref{fig:3Msun}.
As shown in Fig.\,\ref{fig:t-LFUV}, the $\LFUV$ of a pre-MS $3\,\Msun$ star should be much lower than that of an MS star, but, in approaching the MS, should suddenly increase by orders of magnitude. This time evolution has a strong impact on the disk lifetime.
We note that Eq.\,(\ref{eq:LFUV}) is adopted, but $\LFUVph$ always dominates; therefore $\LFUV$ is almost constant with time.

The fact that a disk is dispersed mainly by the PE driven by $\LFUVph$ is the same in both cases in Figs.\,\ref{fig:3Msun} and \ref{fig:importance}. However, for a realistic disk evolution model around IM stars, we claim that stellar evolution is one important ingredient.

We note the difference between the results in Fig.\,\ref{fig:importance} and \citet{Gorti+09}: even though the $\LFUVph, \PhiEUV$, and $\LX$ values are the same, the disk lifetimes differ by about 1 order of magnitude (0.2 and 4\,Myr, respectively).
We speculate that the difference probably originates from the absorption of high-energy photons in disk winds from an inner disk. \citet[][see their section 2.4.1]{Gorti+09} considered this effect, whereas we do not. This effect can suppress the PE rate in the early phase. We will discuss this issue in Sect.\,\ref{sec:caveats-PE}. Nevertheless, our claim that the time-dependent $\LFUVph$ is important for disk evolution is still valid.

\subsection{Disk Lifetime}\label{sec:tdisk}

\begin{figure}[!t]
  \begin{center}
    \includegraphics[width=\hsize,keepaspectratio]{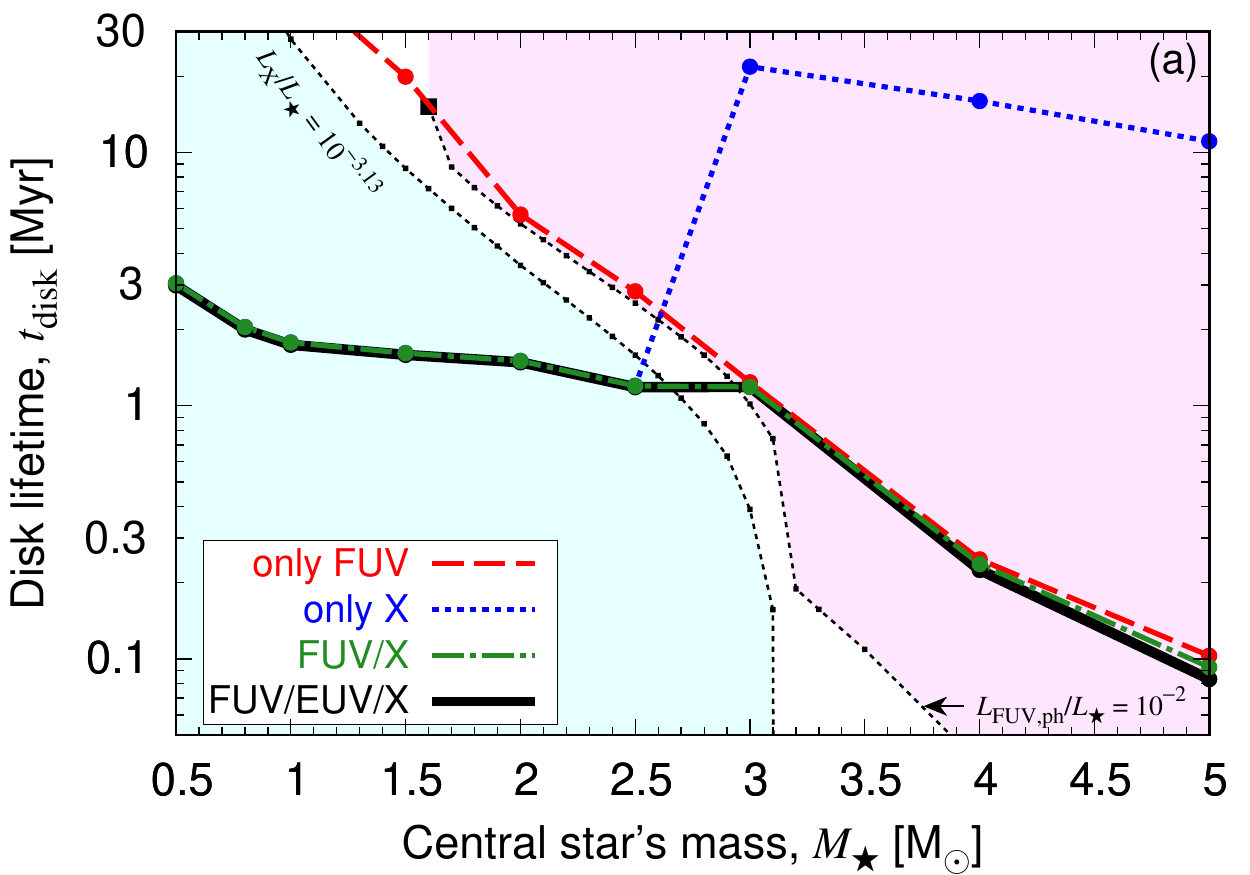}
    \includegraphics[width=\hsize,keepaspectratio]{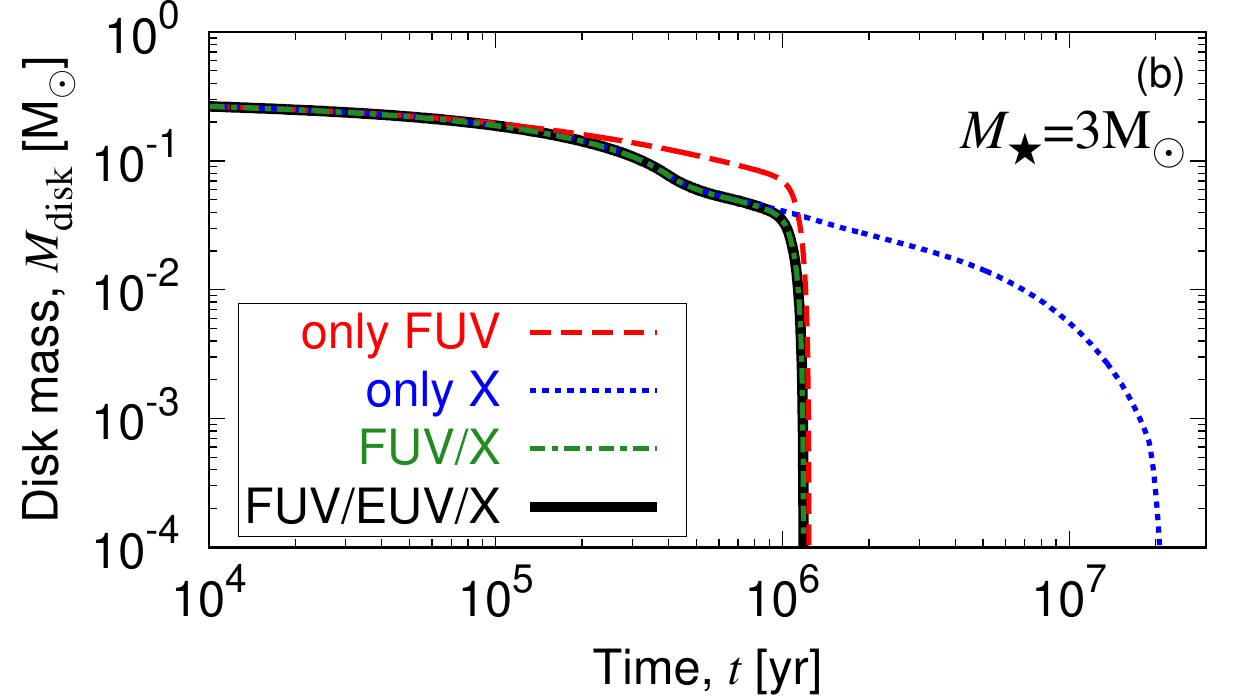}
	\caption{\small{
	Top panel: disk lifetime as a function of $\Mstar$ in the cases with the FUV, EUV, and X-ray PE (black solid line), with only the FUV PE (red dashed line), with only the X-ray PE (blue dotted line) and with the FUV and X-ray PE (green dotted-dashed line).
	The two thin black dotted lines show the time when $\Ro = \Rosat$ (left; i.e., $\LX/\Lstar = 10^{-3.13}$) and $\Teff = 7342\,\K$ (right; i.e., $\LFUVph/\Lstar=10^{-2}$).
	The cyan shaded region illustrates the phase in which the X-ray PE dominates, whereas in the magenta shaded region, the FUV PE dominates.
	Bottom panel: time evolution of disk mass, $\Md$, around a $3\,\Msun$ star.
}}\label{fig:disk_lifetime}
    \end{center}
\end{figure}

We perform a suite of disk evolution simulations around 0.5--$5\,\Msun$ stars as in Sect.\,\ref{sec:overview}. We find that $\tdisk$ decreases with increasing $\Mstar$ (Fig.\,\ref{fig:disk_lifetime}a).

To understand which mechanism plays the dominant role, we also perform three sets of simulations: (i) with only the FUV PE, (ii) with only the X-ray PE, and (iii) without the EUV PE.
The other settings are the same as the fiducial runs (see Table\,\ref{tab:input}).
In the models where we do not include the FUV and EUV PE (the ``only X'' model in Fig.\,\ref{fig:disk_lifetime}), the disk lifetime around $\geq3\,\Msun$ stars increases significantly, while any combination of mechanisms that includes FUV causes similarly short lifetimes for $\geq3\,\Msun$ stars.
These results clearly illustrate that disks around $\geq3\,\Msun$ stars are dispersed mainly by the FUV PE.

Figure\,\ref{fig:disk_lifetime}b shows the evolution of $\Md$ around a $3\,\Msun$ star. 
After the X-ray PE becomes less effective at 0.4\,Myr, it takes time for the FUV PE to become strong at 1.0\,Myr, and then the disks quickly disperse if the FUV PE is considered.

Figure\,\ref{fig:disk_lifetime}a also shows the time when stars reach $\Ro=\Rosat$ (i.e., $\LX=10^{-3.13}\,\Lstar$; the maximum value of $\LX$) and $\Teff=7342\,\rm{K}$ (i.e., $\LFUVph=10^{-2}\,\Lstar$; as an indicative timescale for $\MdotFUV$ to increase).
These timescales decrease with $\Mstar$. This is because higher-mass stars have a shorter KH timescale $\tKH$ (see Eq.\,\ref{eq:tauKH}) and therefore develop a radiative core and have a hotter photosphere more rapidly. 
We note that the $\Teff$ of stars with less than $1.6\,\Msun$ never reaches 7342\,K in the pre-MS and main-sequence (MS hereafter) phases; therefore $\LFUVph$ is always below $10^{-2}\Lstar$.

In the cases with FUV, the disks around $\geq2\,\Msun$ stars disperse after $\LFUVph$ reaches $10^{-2}\,\Lstar$. In the case with only the FUV PE, the disk lifetime around $\sim1.5$--3\,$\Msun$ stars is almost the same as the timescale to reach $\LFUVph=10^{-2}\,\Lstar$.
Therefore, if the X-ray PE is less effective, the disk lifetime around IM stars is determined by the stellar evolution.
We note that even though the $\LFUV$ of 4 and 5$\,\Msun$ stars becomes luminous in the early phase, it takes time for $\Mdotacc$ to decrease and for the disks to disperse.
On the other hand, the disk lifetime around $\lesssim1\,\Msun$ stars in the case with only FUV exceeds 30\,Myr. 
This is because the $\LFUV$ of low-mass stars is dominated by $\LFUVacc$, which is self-regulated; $\LFUVacc$ decreases along with decreasing $\Mdotacc$ over time. Therefore, the PE  mainly by $\LFUVacc$ does not open a gap.

If we compare the cases with and without the X-ray PE, one finds that the disks around $\lesssim2.5\,\Msun$ stars disperse mainly by the X-ray PE.
The influence of the EUV PE on $\tdisk$ is negligible in the entire mass range.
Therefore, under the current settings, $\gtrsim3\,\Msun$ stars disperse mainly by the FUV PE, whereas $\lesssim2.5\,\Msun$ stars disperse by the X-ray PE.
However, we note that, although we adopt the X-ray and EUV PE rates from the literature in this study, they are still under debate (see Sect.\,\ref{sec:caveats-PE}). If our X-ray PE rate is overestimated, then the realistic $\tdisk$ should be in between the $\tdisk$ of the fiducial case and that of the ``only FUV'' case. 
Nevertheless, the importance of the $\LFUVph$ evolution around IM stars is not affected by the uncertainty of the X-ray PE model.

On the high-mass side ($\gtrsim3\,\Msun$), $\tdisk$ decreases with $\Mstar$ because of the shorter $\tKH$, as described above.
Here we explain why we obtain the same trend on the low-mass side. The $\tdisk$ value is almost the same as the timescale of the gap opening, which occurs when $\Mdotacc$ decreases down to $\MdotPE$ (see Sect.\,\ref{sec:overview}).
Both have a similar dependence on $\Mstar$. We chose the input parameter $\alpha$ to reproduce the observed relation $\Mdotacc\propto \Mstar^2$ (Sect.\,\ref{sec:num}).
Around low-mass stars, the X-ray PE dominates, and therefore $\MdotPE\simeq\MdotX$. We adopt the X-ray PE model based on \citet{Owen+12}, which is in proportion to $\LX$.
Figure\,\ref{fig:t-LX} shows that $\LX$ is roughly proportional to $\Mstar^{1.6}$ in the case of 1 Myr old low-mass stars.
Since both $\Mdotacc$ and $\MdotPE$ have a similar correlation with $\Mstar$, the gap-opening timescale is determined by the timescale for $\Mdotacc$ to decrease, that is, the viscous timescale $\tvis$ \citep{Clarke+01}.
Given that $\nu\sub{vis} \propto \Mstar$ (see Sect.\,\ref{sec:num}) and that we assume $R_1$ does not correlate with $\Mstar$,  $\tvis\propto \Mstar^{-1}$.
Therefore, $\Mdotacc$ decreases faster around higher-mass stars and $\tdisk$ decreases with $\Mstar$.
We note that for this correlation, Eq.\,\ref{eq:alpha} is essentially important because this gives the relation $\nu\sub{vis} \propto \Mstar$ (see discussions in  Sect.\,\ref{sec:alp}).

\section{Discussion}\label{sec:discussion}

\subsection{Comparison with Observations}
\label{sec:NIRobs}\label{sec:multi}

In this subsection, we compare our results with observations.
Here we focus only on the gas disk lifetime (see Sect.\,\ref{sec:intro} for dust disk lifetimes); recent H$\alpha$ observations have revealed that the gas disk lifetime around IM stars is shorter than that of low-mass stars \citep{Kennedy+Kenyon09,Yasui+14}.
This is consistent with our results of $\tdisk$ in Fig.\,\ref{fig:disk_lifetime}a.
We again stress that the realistic $\LFUVph$ model with stellar evolution is crucial for this trend on the high-mass side, whereas the $\LX$ and $\alpha$ models are important on the low-mass side.
Since we have not explored the dependence on the input parameters and the PE models are still under debate, we limit ourselves to focusing only on the qualitative results in this study.
We leave the quantitatively detailed discussions for future studies.

\subsection{Dependence on the Variety in X-Ray Luminosity}\label{sec:Prot}
\label{sec:LX}

\begin{figure*}[!t]
  \begin{center}
    \includegraphics[width=0.48\hsize,keepaspectratio]{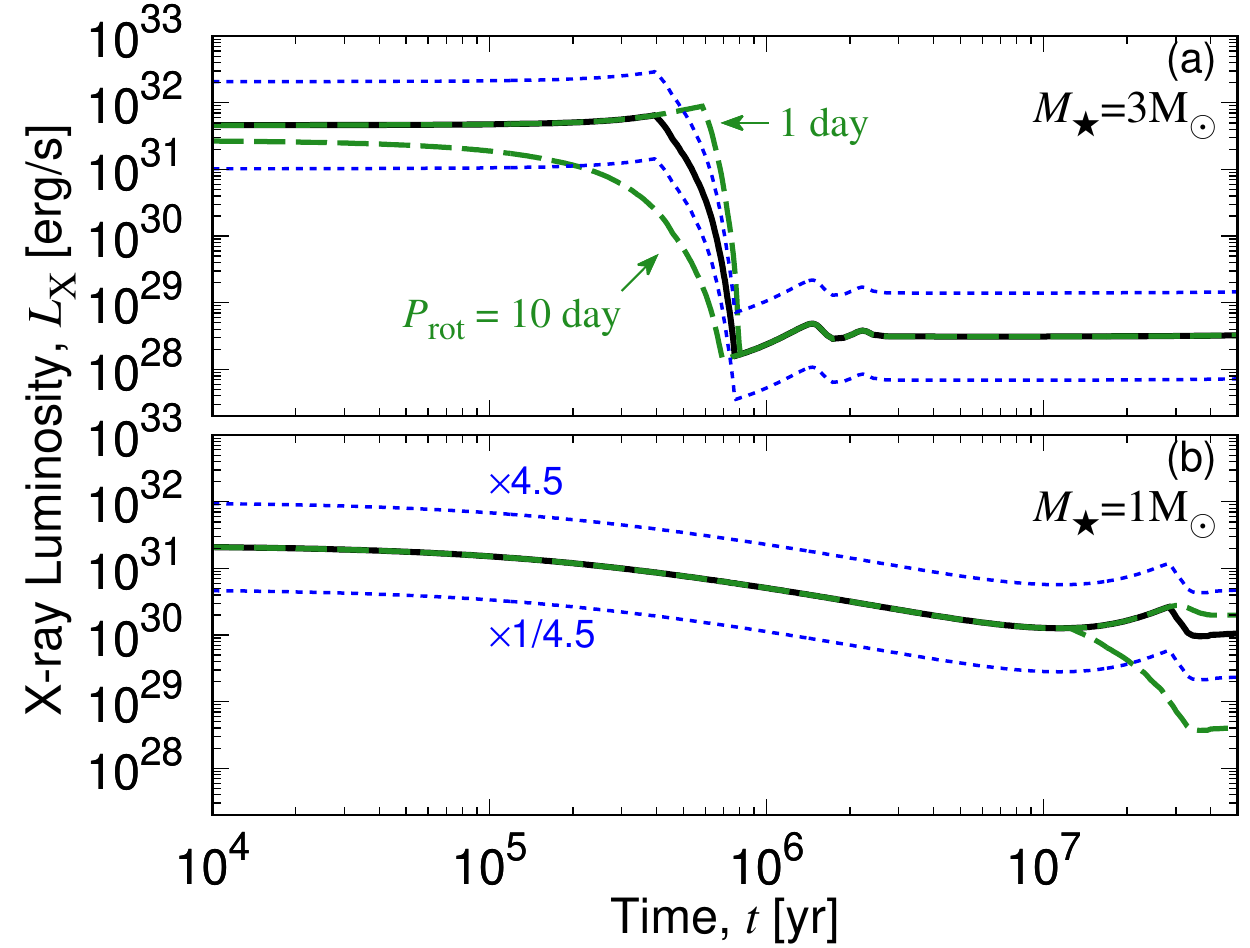}
    \includegraphics[width=0.48\hsize,keepaspectratio]{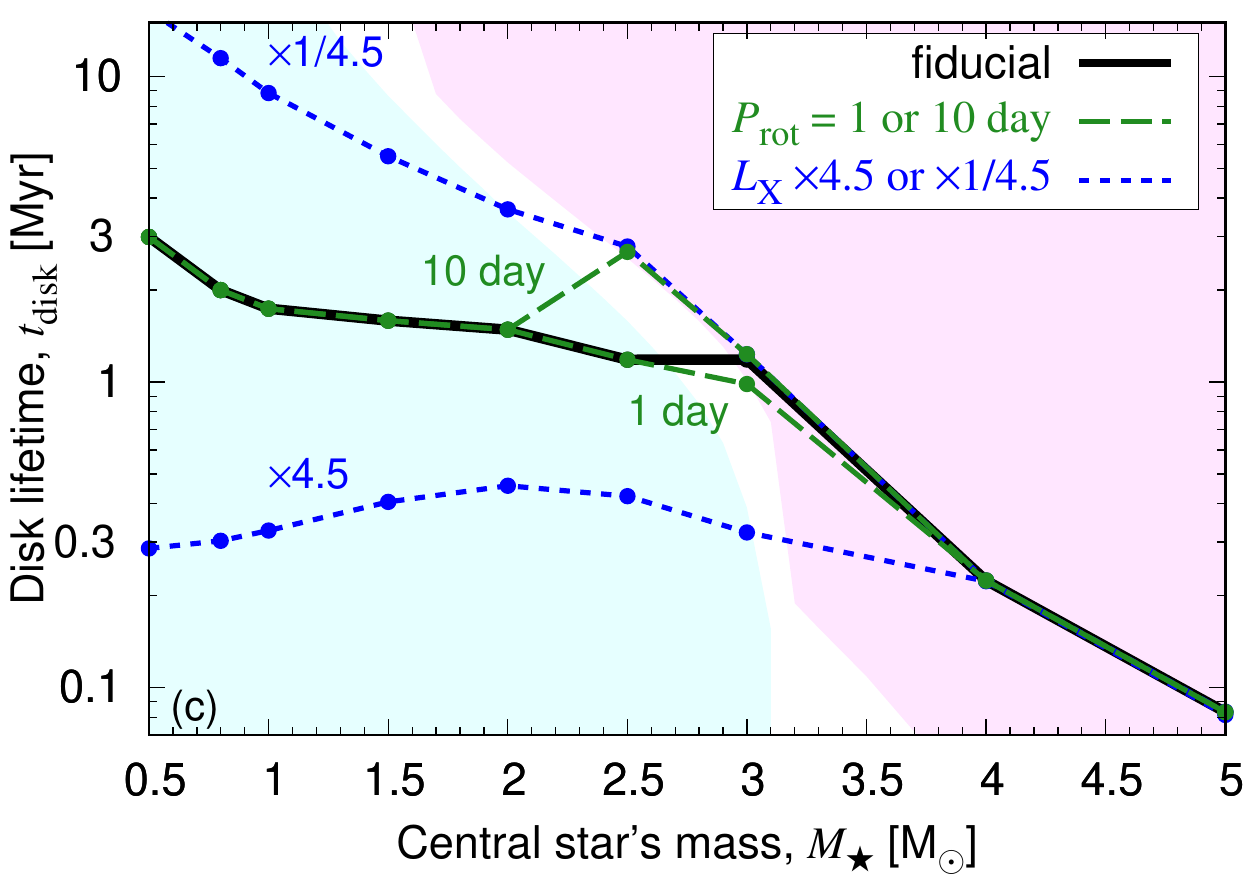}
	\caption{\small{
	(Left panel) Temporal evolution of the $\LX$ of a 3 (top) and 1 (bottom) $\Msun$ star. 
	The black solid lines show the fiducial one (i.e., $\Prot=3\,\rm days$). 
	The dashed green lines show the cases with $\Prot=1$ (top) and 10 (bottom) days.
	The dotted blue lines are $\LX\times 4.5$ (top) and $\LX/4.5$ (bottom), highlighting the observed scatter.
	(Right panel) The $\tdisk$ with the different $\LX$ models.
	The cyan and magenta shaded regions are the same as in Fig.\,\ref{fig:disk_lifetime}a.
	We note that the cyan region depends on $\Prot$ (see the left panel).
}}\label{fig:vary-LX}
    \end{center}
\end{figure*}

We have found that disks around $\lesssim 2.5\,\Msun$ stars are dispersed mainly by the X-ray PE, and therefore $\tdisk$ depends on $\LX$.
Observations have revealed that the stellar $\LX$ has a large variety.
Although in this paper, we have adopted the empirical relation of \citet[][see Eq.\,\ref{eq:LX}]{Wright+11}, the observed data of $\RX$ \citep[see, e.g.,][]{Preibisch+05} exhibit a variety by a factor of 4.5 ($=0.65$\,dex).
Moreover, although we have assumed $\Prot=3$\,days, the observed rotational period of pre-MS stars has a variety from $\sim1$ to 10\,days (see Sect.\,\ref{sec:LXevol}).
In this section, we explore the influence of these varieties on the results of $\tdisk$.

Figures\,\ref{fig:vary-LX}a and \ref{fig:vary-LX}b show the $\LX$ evolution of 3 and $1\,\Msun$ stars, respectively. We consider the cases with $\Prot=1$ and 10\,days and $\LX$ multiplied or divided by a factor of 4.5.
We find that $1\,\Msun$ stars develop a radiative core at $\simeq10\,$Myr, and until then, pre-MS stars are in the saturated regime irrespective of $\Prot$, whereas it happens for $3\,\Msun$ stars in the early ($\simeq0.4$\,Myr) phase.
We note that \citet{Tu+15} claimed that the $\LX$ of $1\,\Msun$ MS stars has a large variety depending on the $\Prot$.
This is because the $\tau\sub{conv}$ of $1\,\Msun$ MS stars is short enough for their $\LX$ to depend on $\Prot$ (see also Eq.\,\ref{eq:LX}). However, our results show that the $\LX$ of pre-MS $1\,\Msun$ stars does not depend on $\Prot$ until $\simeq10$\,Myr.

Figure\,\ref{fig:vary-LX}c shows $\tdisk$ with different $\LX$ models.
Here we adopt fiducial settings other than $\LX$.
We find that the variation in $\Prot$ has little impact on $\tdisk$. On the other hand, if we change $\LX$ by a factor of 4.5, $\tdisk$ changes by up to 1\,dex. The variation of $\LX$ has a larger impact on $\tdisk$ around lower-mass stars.
Therefore, for the detailed comparison with observed disk fractions with time, we need to consider the $\LX$ variation as claimed by \citet{Kimura+16}.

The trend of $\tdisk$ with $\Mstar$ depends on different $\LX$ models:
$\tdisk$ decreases with increasing $\Mstar$ in the low-$\LX$ case, 
whereas the $\tdisk$ of $<3\,\Msun$ stars is almost constant in the high-$\LX$ case.
For the former, the reason is the same as the fiducial case (i.e., the shorter $\tvis$; see Sect.\,\ref{sec:tdisk}).
For the latter, $\MdotX\gg\Mdotacc$ from the beginning, and therefore the gap-opening timescale ($\sim\tdisk$) is determined by $\tauPE(R\sub{gap})=\Sigma/\SigdotPE$, where  $R\sub{gap}$ is the radius where the PE opens a gap.
Below, we briefly show that $\tauPE(R\sub{gap})$ is insensitive to $\Mstar$.
First, $R\sub{gap}\propto \Mstar$ because the location of the peak of $\SigdotX$ is proportional to $\Mstar$ (see Sect.\,\ref{sec:SigdotPE}).
Since we assume $\Sigma\propto R^{-1}\Mstar$ as an initial condition, the initial $\Sigma$ at $R\sub{gap}$ does not depend on $\Mstar$.
Second, $\SigdotPE(R\sub{gap})\simeq\SigdotX\propto \LX R\sub{gap}^{-2}$ (see Eq.\,\ref{eq:SigdotX}), where $\LX\propto \Mstar^{1.6}$ but $R\sub{gap}^{-2}\propto \Mstar^{-2}$. These two opposite effects make the peak $\SigdotX$ value almost constant with $\Mstar$.
Therefore, the $\tauPE$ (and thus $\tdisk$) of $<3\,\Msun$ stars is insensitive to $\Mstar$ in the high-$\LX$ case.

\subsection{Dependence on the Variety in Viscosity}\label{sec:alp}

We have adopted $\alpha\propto\Mstar$ to reproduce the observed relation ($\Mdotacc\propto\Mstar^2$; see Sect.\,\ref{sec:num}), but the physical origin of this relation is still unclear.
In addition, the absolute value of $\alpha$ is also under debate.
As a fiducial value, we adopt a relatively large $\alpha$ value ($=10^{-2}\,(\Mstar/\Msun)$) assuming that the disks are turbulent. However, recent observations \citep[e.g.,][]{Pinte+16,Flaherty+17} and theoretical studies \citep[see, e.g., ][and references therein]{Turner+14} have suggested a low $\alpha$ (e.g., $\lesssim10^{-3}$ from the observations).

\begin{figure}[!t]
  \begin{center}
    \includegraphics[width=\hsize,keepaspectratio]{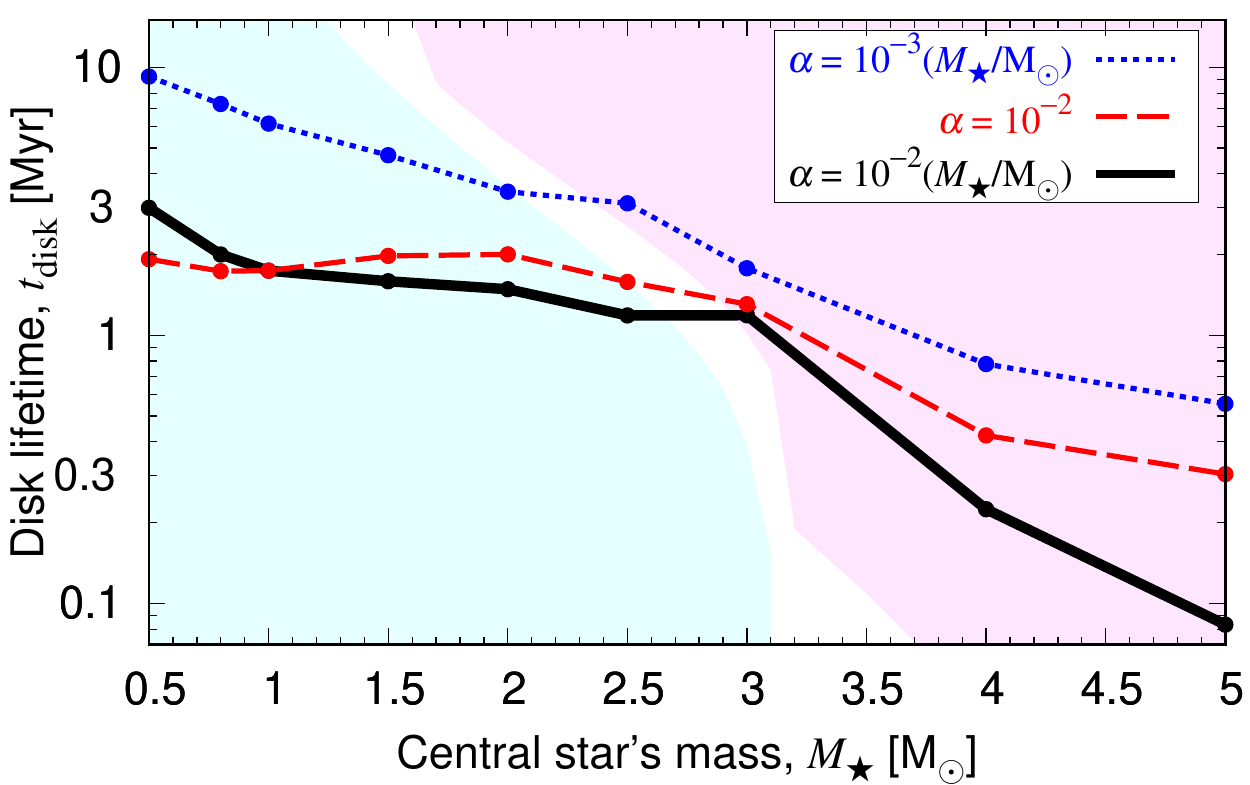}
	\caption{\small{
	Disk lifetime, $\tdisk$, with different $\alpha$ models: $\alpha=10^{-2}(\Mstar/\Msun)$ (black solid; fiducial), $\alpha=10^{-2}$ (red dashed), and $\alpha=10^{-3}(\Mstar/\Msun)$ (blue dotted). 
	The cyan and magenta shaded regions are the same as in Fig.\,\ref{fig:disk_lifetime}a.
}}\label{fig:alp}
    \end{center}
\end{figure}

To explore the dependence of $\tdisk$ on the $\alpha$ model, we simulate disk evolutions with $\alpha=10^{-2}$ (i.e., constant $\alpha$ with $\Mstar$) and $\alpha=10^{-3}\,(\Mstar/\Msun)$ (i.e., 10 times lower than the fiducial model).
Figure\,\ref{fig:alp} shows that the decreasing $\tdisk$ with $\Mstar$ on the high-mass side ($\geq 3\,\Msun$) remains even if we adopt a different $\alpha$ model because the rapid increase of $\LFUVph$ has a dominant role.

We note that the variety in $\alpha$ affects the $\tdisk$ values; a lower $\alpha$ by a factor of 10 results in a larger $\tdisk$ by a factor of $\simeq3$ \citep[as shown in figure\,11 of ][]{Gorti+09}.
We also note that if $\alpha$ is constant with $\Mstar$, the $\tdisk$ value is also constant with $\Mstar$ ($\simeq2$\,Myr) in the range $\Mstar \leq3\,\Msun$. 
Therefore, to compare theoretical $\tdisk$ values with observations, it is crucial to understand the origin of the relation $\Mdotacc\propto\Mstar^2$ and constrain the absolute value of $\alpha$ in protoplanetary disks.

\subsection{Model caveats}
\label{sec:caveats}

In this subsection, we describe the caveats on the PE models, evolution of dust disks, magnetohydrodynamic (MHD) winds, and variations of input parameters.

\label{sec:caveats-PE}

We point out two issues on the PE models.
First, although we adopt the X-ray PE model by \citet{Owen+12}, 
their $\MdotX$ is higher than that of recent RHD simulations with a self-consistent thermochemistry by \citet{Wang+Goodman17} and \citet{Nakatani+18b}.
Therefore, although our results suggest that the disks around $\lesssim2.5\,\Msun$ stars disperse mainly by the X-ray PE (Sect.\,\ref{sec:overview}), the $\tdisk$ of $\lesssim2.5\,\Msun$ stars may be underestimated.
Future works should investigate the long-term disk evolution with the updated X-ray PE rate.
Second, the PE may be suppressed in particular in the early phase in the outer region due to the absorption of high-energy photons.
These photons can be shielded by dense gas, such as accretion flows onto the star \citep{Alexander+04a}, inner disk winds \citep{Bai17,Takasao+18}, stellar winds \citep{Hollenbach+00} and dust grains in the disk atmosphere \citep{Nakatani+18a}.
If the high-energy photons are shielded, the PE rate can decrease by orders of magnitude, and the PE profiles can also be changed (see also Sect.\,\ref{sec:importance}).

There are two issues in the luminosity and spectra of stellar high-energy photons.
First, in this paper, we have used a simple model of $\PhiEUV$, but this is quite uncertain (Sect.\,\ref{sec:PhiEUV}). \citet{Bouret+Catala98} suggested that Herbig Ae/Be stars have $\PhiEUV\sim10^{43}$--$10^{45}$ using an indirect estimation. 
Although the EUV PE has a marginal effect on the disk evolution in our results, we expect that future works constrain the $\PhiEUV$ of IM stars more precisely.
Second, the hardness of the X-ray spectra of young stars remains a matter of debate.
Some observations have suggested that the X-ray spectra of accreting stars may be softer \citep[e.g., ][]{Kastner+02,Kastner+04}. 
\citet{Gorti+09} showed that a softer X-ray spectrum results in a larger PE rate even with the same $\LX$ (see their figure 9). Future studies should investigate the influence of the evolution of the X-ray hardness on the disk evolution.

The uncertainties and varieties in the PE models above would be important for some observational results.
Although most IM stars have a shorter inner disk lifetime (see Sect.\,\ref{sec:NIRobs}), some have a long disk lifetime \citep[e.g., ][]{Panic+08, Fedele+17,Booth+19,Miley+19,Muro-Arena+20}.
These long-lived disks may have a lower PE rate.
Since most of these long-lived disks around Herbig stars are well studied due to the relative ease of detecting their large bright disks, there are a lot of existing high-quality data for theoretical models to be compared with.
Theoretical models should be compared in detail with and explain these observations in future.

\label{sec:caveats-dust}

We stress the importance of the dust disk evolution, which is not considered in this paper.
Previous studies have found that gas and dust disk lifetimes can differ \citep[see, e.g., ][]{Takeuchi+05, Alexander+Armitage07, Gorti+15,Owen+Kollmeier19}.
Since IR observations trace the small dust grains, we need to simulate the long-term evolution of gas and dust to compare theoretical models with IR observations.
The number of dust grains in the disk atmosphere may also affect the FUV PE rate \citep{Gorti+15,Nakatani+20}.
However, the motion and evolution of dust grains are quite complicated; we need to consider a number of effects, such as radial drift \citep{Adachi+76}, gas pressure gradient \citep{Taki+16,Taki+20}, coagulation, fragmentation and collisional cascade \citep{Kobayashi+Tanaka10}, the entrainment in the PE or MHD disk winds \citep{Gorti+15,Miyake+16, Franz+20}.
Future studies with the dust evolution and stellar evolution around IM stars are needed to investigate the realistic lifetimes of dust disks.

\label{sec:caveats-MHD}

In this paper, we have not included MHD disk winds, but recently much attention has been paid to them \citep[e.g.,][]{Suzuki+Inutsuka09,Fromang+13,Lesur+13,Bai+Stone13a,Bai17,Wang+19}.
The MHD winds carry away not only mass but also angular momentum \citep[so-called wind-driven accretion;][]{Bai+Stone13b,Bai16,Suzuki+16}.
\citet{Kunitomo+20} claimed that the MHD and PE winds have different roles \citep[see also recent radiation-MHD simulations by][]{Wang+19,Rodenkirch+20,Gressel+20} and both winds and the wind-driven accretion should be considered for a realistic disk evolution, in particular for disks with weak turbulence.
We will investigate the long-term disk evolution around IM stars including both winds in our next paper.

\label{sec:caveats-inputs}

We have not varied the input parameters in this paper.
The variety of the initial disk condition, $\Mdini$ and $R_1$, should be related to the properties of parental clouds using a disk formation model \citep{Takahashi+13,Kimura+16}. 
For a detailed comparison with the observations of disk fractions over time, we need Monte Carlo simulations covering the variety of input parameters \citep[$\Mdini$ $R_1$, and $\alpha$;][]{Alexander+Armitage09,Kimura+16}.

\label{sec:caveats-star}

Finally, we discuss the variety of stellar evolution.
Although in this paper, we adopted the birthline based on the standard star formation scenario, recent studies have shown that the luminosity of the birthline depends on star formation processes \citep[such as the variety in the entropy of accreting materials or deuterium abundance; see][]{BCG09,Hosokawa+11,Tognelli+15,Kunitomo+17,Kuffmeier+18}.
Stellar $\Teff$ depends on the metallicity and mixing-length parameter $\alpha\sub{MLT}$; a lower metallicity or larger $\alpha\sub{MLT}$ results in a higher $\Teff$ \citep[][]{Kippenhahn+Weigert90}.
Although in this paper we have adopted the solar metallicity and $\alpha\sub{MLT}=2.0$,\footnote{Although \citet{Kunitomo+11} described that $\alpha\sub{MLT}=1.5$, this was a typo.
In standard solar models \citep[see, e.g.,][]{Serenelli+09}, $\alpha\sub{MLT}\simeq2.0$ is suggested.
} the varieties of these parameters affect the $\Teff$ evolution and therefore the $\LFUVph$ and $\PhiEUVph$ evolution.

\subsection{Implications for Planet Formation}
\label{sec:IMplanets}

The disk evolution models have important implications for plant formation.
Since planets form and evolve in a protoplanetary disk, their characteristics may reflect the disk properties. For example, the orbital configuration of planets around IM stars is different from low-mass stars; there is a paucity of close-in planets around $\gtrsim2\,\Msun$ stars \citep[e.g.,][]{Sato+08}. One possible origin is the different disk evolution; the rapid disk dispersal may hinder planets from migrating inward \citep[e.g.,][]{Burkert+Ida07,Currie09,Kunitomo+11}.
Radial velocity surveys have revealed that the occurrence rate of detected giant planets depends upon $\Mstar$ \citep[e.g.,][]{Johnson+10, Reffert+15}.
The mass fraction and/or composition of planet atmospheres can give an indication as to when or where the planet was formed in a disk \citep[][]{Guillot+Hueso06,Ogihara+20,Miley+21}.
We expect that our disk evolution models also lead to the understanding of planet formation processes around IM stars.

\section{SUMMARY AND CONCLUSIONS}\label{sec:conclusion}

We investigated the long-term disk evolution around 0.5--$5\,\Msun$ stars by considering the viscous accretion; the PE mass loss by stellar FUV, EUV, and X-rays; and stellar evolution.
We started calculations from the early phase and initial conditions with a compact ($R_1=50\,$au) and massive ($\Mdini=0.1\,\Mstar$) disk.

We found that the nature of the emission of stellar high-energy photons changes with time;
low-mass stars strongly emit X-rays until the typical disk lifetime (i.e., several Myr), whereas the X-ray luminosity of higher-mass stars decreases and instead, their FUV luminosity rapidly increases due to stellar evolution (e.g., at around $1\,$Myr in the case of $3\,\Msun$ stars).
The critical mass is $\sim2.5\,\Msun$ because the KH timescale becomes comparable to the disk dispersal timescale.
Therefore, the effect of stellar evolution is not negligible, as assumed in previous works, and should be considered for realistic disk evolution models around IM stars.

Our results show that if we consider all of the PE mechanisms (X-ray, EUV, and FUV) with stellar evolution, then $\tdisk$ decreases with $\Mstar$.
The same trend has also been suggested by H$\alpha$ observations. For a detailed comparison with the observations, our models should be refined in future work.
Although we have adopted the PE models from the literature, they have recently been revisited with a self-consistent thermochemistry.
Our models simulate the evolution of gas disks, but the evolution of dust disks is crucially important for the comparison with IR observations.
We considered viscous accretion and PE, but other physical processes, such as MHD disk winds or magnetic braking, should also be considered simultaneously.
We have not surveyed large parameter ranges of $R_1$, $\Mdini$, and $\alpha$. 

The evolution models of protoplanetary disks are crucially important for planet formation theory. 
We expect that the disk evolution models presented in this paper will lead to the understanding of planet formation around IM stars.

\acknowledgments

We are grateful for the simulation results of star formation provided by Dr. Steven W. Stahler.
We are also grateful to Drs. Kei E. I. Tanaka, Chikako Yasui, Masahiro Ikoma, Taishi Nakamoto, Hideko Nomura,
Philip J. Armitage, Richard D. Alexander, Jaehan Bae, Kenji Hamaguchi, Shinsuke Takasao, 
Shu-ichiro Inutsuka, and Hiroshi Kobayashi
for fruitful discussions and comments.
We appreciate the constructive comments of the anonymous referee, which helped us to improve this paper.
M.K. and S.I. thank the University of Leeds for the financial support through the International Mobility Fund, and hospitality during their stay in Leeds.
This work was supported by JSPS KAKENHI grant Nos. 12J09296, 23244027, 15H02065, 16H02160, 17H01105, 17H01153, and 20K14542. The work of O.P. is funded by the Royal Society Dorothy Hodgkin Fellowship. J.M.M. is supported through the University of Leeds Doctoral Scholarship.
%
%
%
%
%
\software{MESA \citep[version 2258;][]{Paxton+11},
          Cloudy \citep[version 13.04;][]{Ferland+13},
          Numpy \citep{vanderWalt+11},
          WebPlotDigitizer (version 4.2; \url{https://automeris.io/WebPlotDigitizer})
          }



\appendix


\section{Dependence of photospheric UV luminosity on the stellar surface gravity}\label{sec:Lph-g}

\begin{figure}[!t]
  \begin{center}
    \includegraphics[width=0.48\hsize,keepaspectratio]{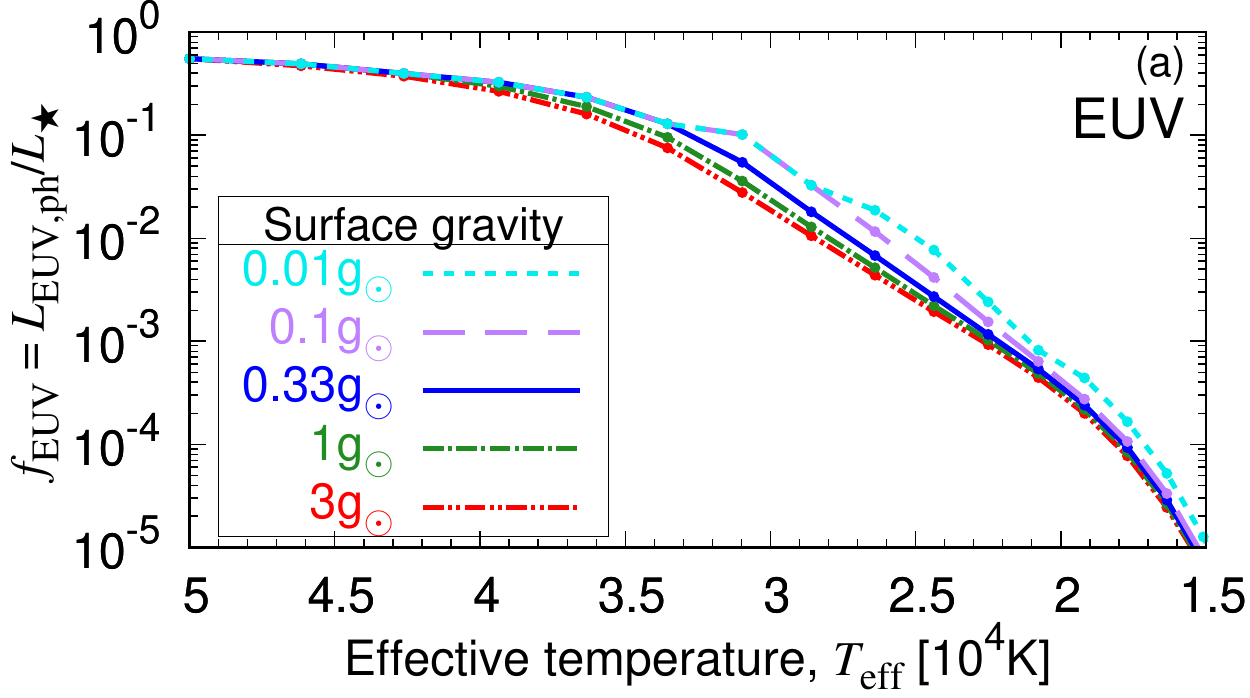}
    \includegraphics[width=0.48\hsize,keepaspectratio]{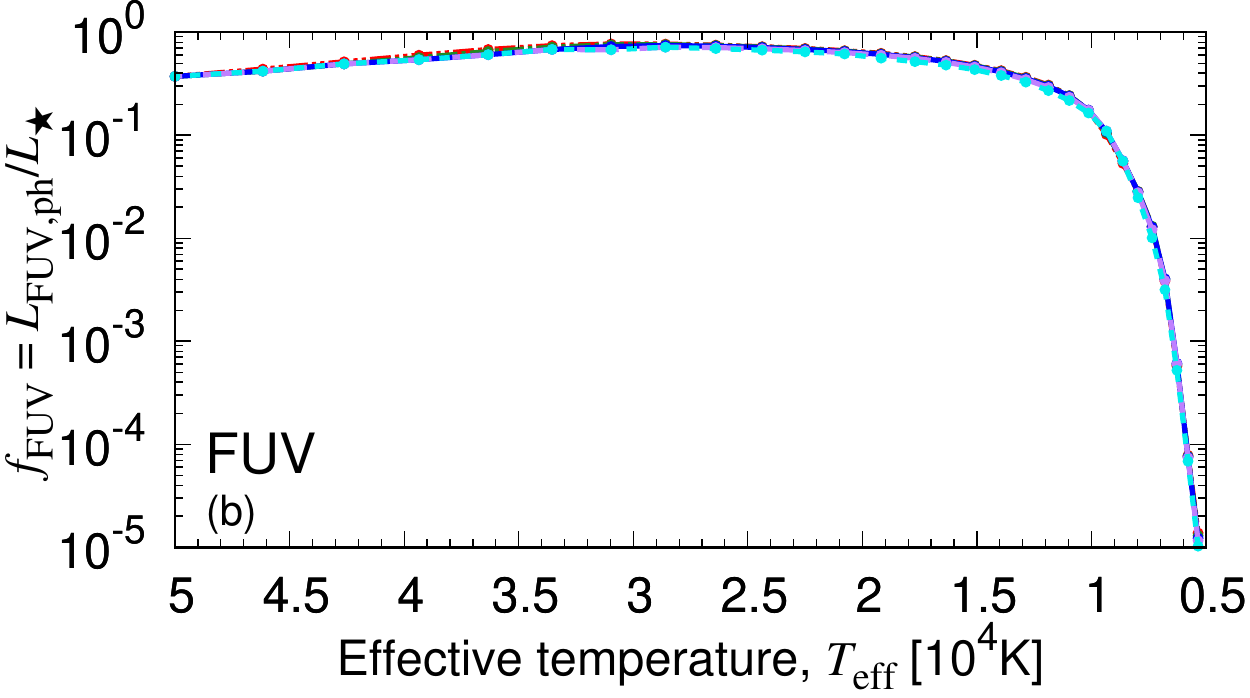}
	\caption{\small{
	$\fredEUV$ (left panel) and $\fredFUV$ (right panel) with varying $g=3$ (red double dotted-dashed lines), 1 (green dotted-dashed lines), 0.33 (fiducial; blue solid lines), 0.1 (purple dashed lines) and 0.01 (cyan dotted lines) $\gsun$.
}}\label{fig:fred-compg}
    \end{center}
\end{figure}

In Sect.\,\ref{sec:spect}, we derived the empirical formulae of photospheric FUV and EUV luminosities in the case of $g=0.33\,\gsun$.
We note that there is a variety in $\log g$ of 0.5--$5\,\Msun$ pre-MS stars; from 0.1 to 10\,Myr, it ranges from 2.7 to 4.3.
Figure\,\ref{fig:fred-compg} shows the weak dependence of $\fredEUV$ and $\fredFUV$ on $\log g$.
We find that the difference of $\fredFUV$ from the fiducial case with $0.33\,\gsun$ is at most 13\%, but that of $\fredEUV$ is up to a factor of 3.
In this paper, we neglect this weak dependence for simplicity.

\section{Disk evolution around low-mass stars}
\label{app:lowmass}

In this Appendix, we show the disk evolution around low-mass stars in our model.
Since the X-ray PE is a matter of debate (see Sect.\,\ref{sec:caveats-PE}), it should be noted that the results may be updated in future work.

\begin{figure*}[!t]
  \begin{center}
    \includegraphics[width=\hsize,keepaspectratio]{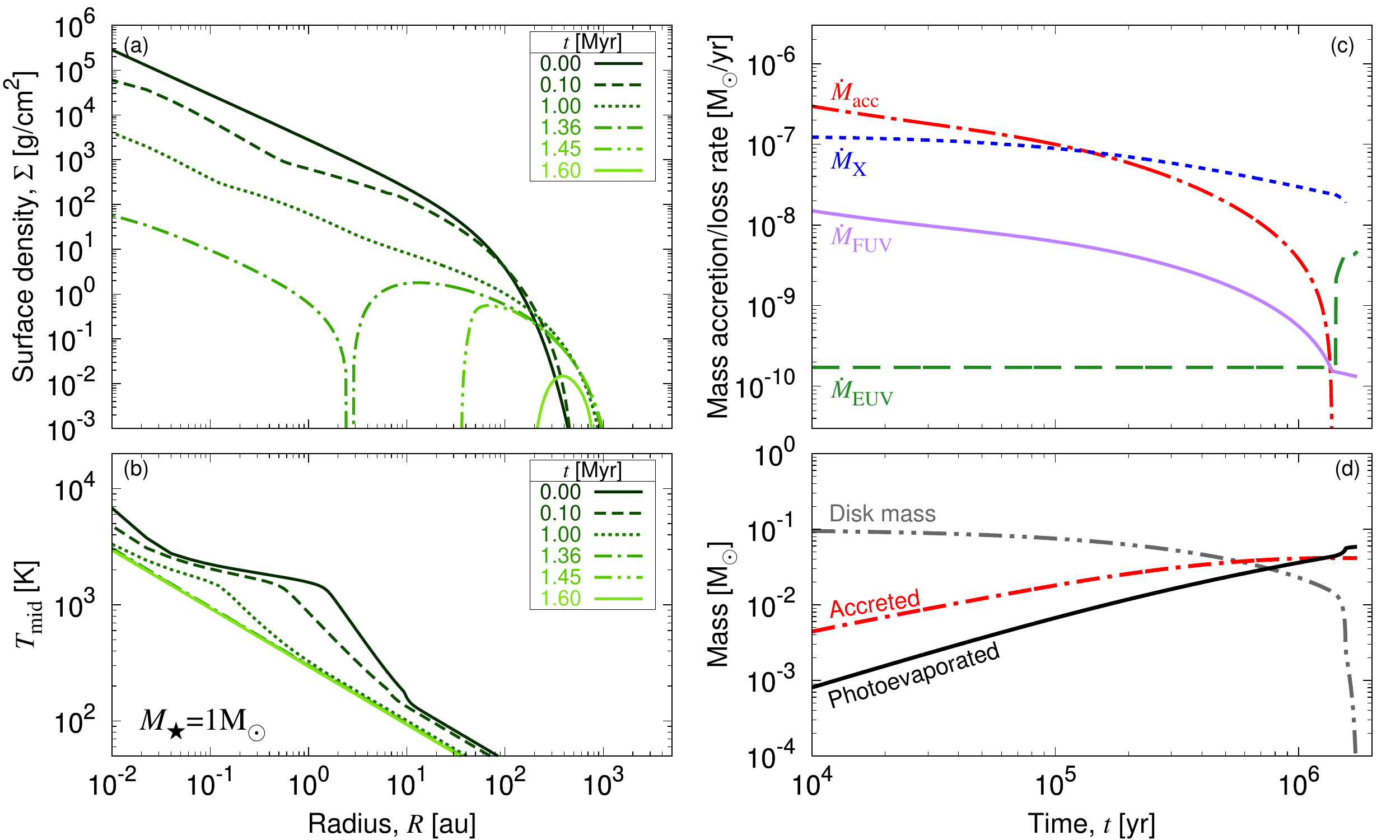}
	\caption{\small{
	Same as Fig.\,\ref{fig:3Msun} but around a $1\,\Msun$ star.
}}\label{fig:1Msun}
    \end{center}
\end{figure*}

Figure\,\ref{fig:1Msun} shows the disk evolution around a $1\,\Msun$ star.
The qualitative behavior of the surface density evolution is the same as the $3\,\Msun$ star case (Sect.\,\ref{sec:overview}).
However, unlike the case of IM stars (Fig.\,\ref{fig:3Msun}), $\MdotX$ is always larger than $\MdotEUV$ and $\MdotFUV$ (see, however, the caveats in Sect.\,\ref{sec:caveats-PE}).
This is because the $\LX$ of $\lesssim 1\,\Msun$ young stars is in the saturated regime and therefore as large as $\sim10^{29}$--$10^{31}\,\rm{erg/s}$.
Therefore, most materials are lost by either accretion or the X-ray PE.

We note that, as described in \citet[][see their section\,4.4]{Kunitomo+20}, we see the gradual decrease of $\MdotX$ over 3\,Myr, but the qualitative behavior described above is the same as the cases with constant $\LX$ in the previous works \citep[e.g.,][]{Owen+10}.
This is expected from the long KH timescale of low-mass stars (see Sect.\,\ref{sec:intro}).

\begin{thebibliography}{}
\expandafter\ifx\csname natexlab\endcsname\relax\def\natexlab#1{#1}\fi
\providecommand{\url}[1]{\href{#1}{#1}}
\providecommand{\dodoi}[1]{doi:~\href{http://doi.org/#1}{\nolinkurl{#1}}}
\providecommand{\doeprint}[1]{\href{http://ascl.net/#1}{\nolinkurl{http://ascl.net/#1}}}
\providecommand{\doarXiv}[1]{\href{https://arxiv.org/abs/#1}{\nolinkurl{https://arxiv.org/abs/#1}}}

\bibitem[{{Adachi} {et~al.}(1976){Adachi}, {Hayashi}, \&
  {Nakazawa}}]{Adachi+76}
{Adachi}, I., {Hayashi}, C., \& {Nakazawa}, K. 1976, Progress of Theoretical
  Physics, 56, 1756, \dodoi{10.1143/PTP.56.1756}

\bibitem[{{Adams} {et~al.}(2004){Adams}, {Hollenbach}, {Laughlin}, \&
  {Gorti}}]{Adams+04}
{Adams}, F.~C., {Hollenbach}, D., {Laughlin}, G., \& {Gorti}, U. 2004, \apj,
  611, 360, \dodoi{10.1086/421989}

\bibitem[{{Alexander} {et~al.}(2014){Alexander}, {Pascucci}, {Andrews},
  {Armitage}, \& {Cieza}}]{Alexander+14}
{Alexander}, R., {Pascucci}, I., {Andrews}, S., {Armitage}, P., \& {Cieza}, L.
  2014, in Protostars and Planets VI, ed. H.~{Beuther}, R.~S. {Klessen}, C.~P.
  {Dullemond}, \& T.~{Henning} (University of Arizona Press), 475--496,
  \dodoi{10.2458/azu_uapress_9780816531240-ch021}

\bibitem[{{Alexander} \& {Armitage}(2007)}]{Alexander+Armitage07}
{Alexander}, R.~D., \& {Armitage}, P.~J. 2007, \mnras, 375, 500,
  \dodoi{10.1111/j.1365-2966.2006.11341.x}

\bibitem[{{Alexander} \& {Armitage}(2009)}]{Alexander+Armitage09}
---. 2009, \apj, 704, 989, \dodoi{10.1088/0004-637X/704/2/989}

\bibitem[{{Alexander} {et~al.}(2004){Alexander}, {Clarke}, \&
  {Pringle}}]{Alexander+04a}
{Alexander}, R.~D., {Clarke}, C.~J., \& {Pringle}, J.~E. 2004, \mnras, 348,
  879, \dodoi{10.1111/j.1365-2966.2004.07401.x}

\bibitem[{{Alexander} {et~al.}(2006{\natexlab{a}}){Alexander}, {Clarke}, \&
  {Pringle}}]{Alexander+06b}
---. 2006{\natexlab{a}}, \mnras, 369, 229,
  \dodoi{10.1111/j.1365-2966.2006.10294.x}

\bibitem[{{Alexander} {et~al.}(2006{\natexlab{b}}){Alexander}, {Clarke}, \&
  {Pringle}}]{Alexander+06a}
---. 2006{\natexlab{b}}, \mnras, 369, 216,
  \dodoi{10.1111/j.1365-2966.2006.10293.x}

\bibitem[{{Andrews} {et~al.}(2013){Andrews}, {Rosenfeld}, {Kraus}, \&
  {Wilner}}]{Andrews+13}
{Andrews}, S.~M., {Rosenfeld}, K.~A., {Kraus}, A.~L., \& {Wilner}, D.~J. 2013,
  \apj, 771, 129, \dodoi{10.1088/0004-637X/771/2/129}

\bibitem[{{Andrews} {et~al.}(2018){Andrews}, {Terrell}, {Tripathi}, {Ansdell},
  {Williams}, \& {Wilner}}]{Andrews+18}
{Andrews}, S.~M., {Terrell}, M., {Tripathi}, A., {et~al.} 2018, \apj, 865, 157,
  \dodoi{10.3847/1538-4357/aadd9f}

\bibitem[{{Andrews} {et~al.}(2010){Andrews}, {Wilner}, {Hughes}, {Qi}, \&
  {Dullemond}}]{Andrews+10}
{Andrews}, S.~M., {Wilner}, D.~J., {Hughes}, A.~M., {Qi}, C., \& {Dullemond},
  C.~P. 2010, \apj, 723, 1241, \dodoi{10.1088/0004-637X/723/2/1241}

\bibitem[{{Ansdell} {et~al.}(2018){Ansdell}, {Williams}, {Trapman}, {van
  Terwisga}, {Facchini}, {Manara}, {van der Marel}, {Miotello}, {Tazzari},
  {Hogerheijde}, {Guidi}, {Testi}, \& {van Dishoeck}}]{Ansdell+18}
{Ansdell}, M., {Williams}, J.~P., {Trapman}, L., {et~al.} 2018, \apj, 859, 21,
  \dodoi{10.3847/1538-4357/aab890}

\bibitem[{{Armitage}(2000)}]{Armitage00}
{Armitage}, P.~J. 2000, \aap, 362, 968.
\newblock \doarXiv{astro-ph/0007044}

\bibitem[{{Bahcall} {et~al.}(2005){Bahcall}, {Basu}, {Pinsonneault}, \&
  {Serenelli}}]{Bahcall+05}
{Bahcall}, J.~N., {Basu}, S., {Pinsonneault}, M., \& {Serenelli}, A.~M. 2005,
  \apj, 618, 1049, \dodoi{10.1086/426070}

\bibitem[{{Bai}(2016)}]{Bai16}
{Bai}, X.-N. 2016, \apj, 821, 80, \dodoi{10.3847/0004-637X/821/2/80}

\bibitem[{{Bai}(2017)}]{Bai17}
---. 2017, \apj, 845, 75, \dodoi{10.3847/1538-4357/aa7dda}

\bibitem[{{Bai} \& {Stone}(2013{\natexlab{a}})}]{Bai+Stone13a}
{Bai}, X.-N., \& {Stone}, J.~M. 2013{\natexlab{a}}, \apj, 767, 30,
  \dodoi{10.1088/0004-637X/767/1/30}

\bibitem[{{Bai} \& {Stone}(2013{\natexlab{b}})}]{Bai+Stone13b}
---. 2013{\natexlab{b}}, \apj, 769, 76, \dodoi{10.1088/0004-637X/769/1/76}

\bibitem[{{Balbus} \& {Hawley}(1991)}]{Balbus+Hawley91}
{Balbus}, S.~A., \& {Hawley}, J.~F. 1991, \apj, 376, 214,
  \dodoi{10.1086/170270}

\bibitem[{{Baraffe} {et~al.}(2009){Baraffe}, {Chabrier}, \& {Gallardo}}]{BCG09}
{Baraffe}, I., {Chabrier}, G., \& {Gallardo}, J. 2009, \apjl, 702, L27,
  \dodoi{10.1088/0004-637X/702/1/L27}

\bibitem[{{Booth} {et~al.}(2019){Booth}, {Walsh}, {Ilee}, {Notsu}, {Qi},
  {Nomura}, \& {Akiyama}}]{Booth+19}
{Booth}, A.~S., {Walsh}, C., {Ilee}, J.~D., {et~al.} 2019, \apjl, 882, L31,
  \dodoi{10.3847/2041-8213/ab3645}

\bibitem[{{Bouret} \& {Catala}(1998)}]{Bouret+Catala98}
{Bouret}, J.-C., \& {Catala}, C. 1998, \aap, 340, 163

\bibitem[{{Bouvier}(2008)}]{Bouvier08}
{Bouvier}, J. 2008, \aap, 489, L53, \dodoi{10.1051/0004-6361:200810574}

\bibitem[{{Burkert} \& {Ida}(2007)}]{Burkert+Ida07}
{Burkert}, A., \& {Ida}, S. 2007, \apj, 660, 845, \dodoi{10.1086/512538}

\bibitem[{{Calvet} \& {Gullbring}(1998)}]{Calvet+Gullbring98}
{Calvet}, N., \& {Gullbring}, E. 1998, \apj, 509, 802, \dodoi{10.1086/306527}

\bibitem[{{Calvet} {et~al.}(2004){Calvet}, {Muzerolle}, {Brice{\~n}o},
  {Hern{\'a}ndez}, {Hartmann}, {Saucedo}, \& {Gordon}}]{Calvet+04}
{Calvet}, N., {Muzerolle}, J., {Brice{\~n}o}, C., {et~al.} 2004, \aj, 128,
  1294, \dodoi{10.1086/422733}

\bibitem[{{Carpenter} {et~al.}(2006){Carpenter}, {Mamajek}, {Hillenbrand}, \&
  {Meyer}}]{Carpenter+06}
{Carpenter}, J.~M., {Mamajek}, E.~E., {Hillenbrand}, L.~A., \& {Meyer}, M.~R.
  2006, \apjl, 651, L49, \dodoi{10.1086/509121}

\bibitem[{{Castelli} \& {Kurucz}(2003)}]{Castelli+Kurucz03}
{Castelli}, F., \& {Kurucz}, R.~L. 2003, in IAU Symposium, Vol. 210, Modelling
  of Stellar Atmospheres, ed. N.~{Piskunov}, W.~W. {Weiss}, \& D.~F. {Gray},
  A20.
\newblock \doarXiv{astro-ph/0405087}

\bibitem[{{Chandrasekhar}(1961)}]{Chandrasekhar61}
{Chandrasekhar}, S. 1961, {Hydrodynamic and hydromagnetic stability}
  (Oxford:Clarendon)

\bibitem[{{Clarke} {et~al.}(2001){Clarke}, {Gendrin}, \&
  {Sotomayor}}]{Clarke+01}
{Clarke}, C.~J., {Gendrin}, A., \& {Sotomayor}, M. 2001, \mnras, 328, 485,
  \dodoi{10.1046/j.1365-8711.2001.04891.x}

\bibitem[{Cox \& Giuli(1968)}]{Cox+Giuli68}
Cox, J., \& Giuli, R. 1968, Gordon and Breach, New York, 401

\bibitem[{{Currie}(2009)}]{Currie09}
{Currie}, T. 2009, \apjl, 694, L171, \dodoi{10.1088/0004-637X/694/2/L171}

\bibitem[{{Ercolano} {et~al.}(2008){Ercolano}, {Drake}, {Raymond}, \&
  {Clarke}}]{Ercolano+08}
{Ercolano}, B., {Drake}, J.~J., {Raymond}, J.~C., \& {Clarke}, C.~C. 2008,
  \apj, 688, 398, \dodoi{10.1086/590490}

\bibitem[{{Ercolano} \& {Pascucci}(2017)}]{Ercolano+Pascucci17}
{Ercolano}, B., \& {Pascucci}, I. 2017, Royal Society Open Science, 4, 170114,
  \dodoi{10.1098/rsos.170114}

\bibitem[{{Fedele} {et~al.}(2017){Fedele}, {Carney}, {Hogerheijde}, {Walsh},
  {Miotello}, {Klaassen}, {Bruderer}, {Henning}, \& {van Dishoeck}}]{Fedele+17}
{Fedele}, D., {Carney}, M., {Hogerheijde}, M.~R., {et~al.} 2017, \aap, 600,
  A72, \dodoi{10.1051/0004-6361/201629860}

\bibitem[{{Ferland} {et~al.}(2013){Ferland}, {Porter}, {van Hoof}, {Williams},
  {Abel}, {Lykins}, {Shaw}, {Henney}, \& {Stancil}}]{Ferland+13}
{Ferland}, G.~J., {Porter}, R.~L., {van Hoof}, P.~A.~M., {et~al.} 2013, \rmxaa,
  49, 137.
\newblock \doarXiv{1302.4485}

\bibitem[{{Flaccomio} {et~al.}(2003){Flaccomio}, {Damiani}, {Micela},
  {Sciortino}, {Harnden}, {Murray}, \& {Wolk}}]{Flaccomio+03}
{Flaccomio}, E., {Damiani}, F., {Micela}, G., {et~al.} 2003, \apj, 582, 398,
  \dodoi{10.1086/344536}

\bibitem[{{Flaherty} {et~al.}(2017){Flaherty}, {Hughes}, {Rose}, {Simon}, {Qi},
  {Andrews}, {K{\'o}sp{\'a}l}, {Wilner}, {Chiang}, {Armitage}, \&
  {Bai}}]{Flaherty+17}
{Flaherty}, K.~M., {Hughes}, A.~M., {Rose}, S.~C., {et~al.} 2017, \apj, 843,
  150, \dodoi{10.3847/1538-4357/aa79f9}

\bibitem[{{Font} {et~al.}(2004){Font}, {McCarthy}, {Johnstone}, \&
  {Ballantyne}}]{Font+04}
{Font}, A.~S., {McCarthy}, I.~G., {Johnstone}, D., \& {Ballantyne}, D.~R. 2004,
  \apj, 607, 890, \dodoi{10.1086/383518}

\bibitem[{{Franz} {et~al.}(2020){Franz}, {Picogna}, {Ercolano}, \&
  {Birnstiel}}]{Franz+20}
{Franz}, R., {Picogna}, G., {Ercolano}, B., \& {Birnstiel}, T. 2020, \aap, 635,
  A53, \dodoi{10.1051/0004-6361/201936615}

\bibitem[{{Fromang} {et~al.}(2013){Fromang}, {Latter}, {Lesur}, \&
  {Ogilvie}}]{Fromang+13}
{Fromang}, S., {Latter}, H., {Lesur}, G., \& {Ogilvie}, G.~I. 2013, \aap, 552,
  A71, \dodoi{10.1051/0004-6361/201220016}

\bibitem[{{Gallet} \& {Bouvier}(2013)}]{Gallet+Bouvier13}
{Gallet}, F., \& {Bouvier}, J. 2013, \aap, 556, A36,
  \dodoi{10.1051/0004-6361/201321302}

\bibitem[{{Gorti} {et~al.}(2009){Gorti}, {Dullemond}, \&
  {Hollenbach}}]{Gorti+09}
{Gorti}, U., {Dullemond}, C.~P., \& {Hollenbach}, D. 2009, \apj, 705, 1237,
  \dodoi{10.1088/0004-637X/705/2/1237}

\bibitem[{{Gorti} \& {Hollenbach}(2009)}]{Gorti+Hollenbach09}
{Gorti}, U., \& {Hollenbach}, D. 2009, \apj, 690, 1539,
  \dodoi{10.1088/0004-637X/690/2/1539}

\bibitem[{{Gorti} {et~al.}(2015){Gorti}, {Hollenbach}, \&
  {Dullemond}}]{Gorti+15}
{Gorti}, U., {Hollenbach}, D., \& {Dullemond}, C.~P. 2015, \apj, 804, 29,
  \dodoi{10.1088/0004-637X/804/1/29}

\bibitem[{{Gorti} {et~al.}(2016){Gorti}, {Liseau}, {S{\'a}ndor}, \&
  {Clarke}}]{Gorti+16}
{Gorti}, U., {Liseau}, R., {S{\'a}ndor}, Z., \& {Clarke}, C. 2016, \ssr, 205,
  125, \dodoi{10.1007/s11214-015-0228-x}

\bibitem[{{Gregory} {et~al.}(2016){Gregory}, {Adams}, \& {Davies}}]{Gregory+16}
{Gregory}, S.~G., {Adams}, F.~C., \& {Davies}, C.~L. 2016, \mnras, 457, 3836,
  \dodoi{10.1093/mnras/stw259}

\bibitem[{{Gressel} {et~al.}(2020){Gressel}, {Ramsey}, {Brinch}, {Nelson},
  {Turner}, \& {Bruderer}}]{Gressel+20}
{Gressel}, O., {Ramsey}, J.~P., {Brinch}, C., {et~al.} 2020, \apj, 896, 126,
  \dodoi{10.3847/1538-4357/ab91b7}

\bibitem[{{G{\"u}del}(2004)}]{Gudel04}
{G{\"u}del}, M. 2004, \aapr, 12, 71, \dodoi{10.1007/s00159-004-0023-2}

\bibitem[{Guillot \& Hueso(2006)}]{Guillot+Hueso06}
Guillot, T., \& Hueso, R. 2006, Monthly Notices of the Royal Astronomical
  Society: Letters, 367, L47, \dodoi{10.1111/j.1745-3933.2006.00137.x}

\bibitem[{{Hamaguchi} {et~al.}(2005){Hamaguchi}, {Yamauchi}, \&
  {Koyama}}]{Hamaguchi+05}
{Hamaguchi}, K., {Yamauchi}, S., \& {Koyama}, K. 2005, \apj, 618, 360,
  \dodoi{10.1086/423192}

\bibitem[{{Hamidouche} {et~al.}(2008){Hamidouche}, {Wang}, \&
  {Looney}}]{Hamidouche+08}
{Hamidouche}, M., {Wang}, S., \& {Looney}, L.~W. 2008, \aj, 135, 1474,
  \dodoi{10.1088/0004-6256/135/4/1474}

\bibitem[{{Haworth} \& {Clarke}(2019)}]{Haworth+Clarke19}
{Haworth}, T.~J., \& {Clarke}, C.~J. 2019, \mnras, 485, 3895,
  \dodoi{10.1093/mnras/stz706}

\bibitem[{{Hayashi}(1961)}]{Hayashi61}
{Hayashi}, C. 1961, \pasj, 13, 450

\bibitem[{{Herbig}(1960)}]{Herbig60}
{Herbig}, G.~H. 1960, \apjs, 4, 337, \dodoi{10.1086/190050}

\bibitem[{{Hern{\'a}ndez} {et~al.}(2005){Hern{\'a}ndez}, {Calvet}, {Hartmann},
  {Brice{\~n}o}, {Sicilia-Aguilar}, \& {Berlind}}]{Hernandez+05}
{Hern{\'a}ndez}, J., {Calvet}, N., {Hartmann}, L., {et~al.} 2005, \aj, 129,
  856, \dodoi{10.1086/426918}

\bibitem[{{Hillenbrand} {et~al.}(1992){Hillenbrand}, {Strom}, {Vrba}, \&
  {Keene}}]{Hillenbrand+92}
{Hillenbrand}, L.~A., {Strom}, S.~E., {Vrba}, F.~J., \& {Keene}, J. 1992, \apj,
  397, 613, \dodoi{10.1086/171819}

\bibitem[{{Hollenbach} {et~al.}(1994){Hollenbach}, {Johnstone}, {Lizano}, \&
  {Shu}}]{Hollenbach+94}
{Hollenbach}, D., {Johnstone}, D., {Lizano}, S., \& {Shu}, F. 1994, \apj, 428,
  654, \dodoi{10.1086/174276}

\bibitem[{{Hollenbach} {et~al.}(2000){Hollenbach}, {Yorke}, \&
  {Johnstone}}]{Hollenbach+00}
{Hollenbach}, D.~J., {Yorke}, H.~W., \& {Johnstone}, D. 2000, {Disk Dispersal
  around Young Stars} (University of Arizona Press), 401

\bibitem[{{Hosokawa} {et~al.}(2011){Hosokawa}, {Offner}, \&
  {Krumholz}}]{Hosokawa+11}
{Hosokawa}, T., {Offner}, S.~S.~R., \& {Krumholz}, M.~R. 2011, \apj, 738, 140,
  \dodoi{10.1088/0004-637X/738/2/140}

\bibitem[{{Huenemoerder} {et~al.}(2009){Huenemoerder}, {Schulz}, {Testa},
  {Kesich}, \& {Canizares}}]{Huenemoerder+09}
{Huenemoerder}, D.~P., {Schulz}, N.~S., {Testa}, P., {Kesich}, A., \&
  {Canizares}, C.~R. 2009, \apj, 707, 942, \dodoi{10.1088/0004-637X/707/2/942}

\bibitem[{{Ingleby} {et~al.}(2011){Ingleby}, {Calvet}, {Hern{\'a}ndez},
  {Brice{\~n}o}, {Espaillat}, {Miller}, {Bergin}, \& {Hartmann}}]{Ingleby+11}
{Ingleby}, L., {Calvet}, N., {Hern{\'a}ndez}, J., {et~al.} 2011, \aj, 141, 127,
  \dodoi{10.1088/0004-6256/141/4/127}

\bibitem[{{Johnson} {et~al.}(2010){Johnson}, {Aller}, {Howard}, \&
  {Crepp}}]{Johnson+10}
{Johnson}, J.~A., {Aller}, K.~M., {Howard}, A.~W., \& {Crepp}, J.~R. 2010,
  \pasp, 122, 905, \dodoi{10.1086/655775}

\bibitem[{{Judge} {et~al.}(2003){Judge}, {Solomon}, \& {Ayres}}]{Judge+03}
{Judge}, P.~G., {Solomon}, S.~C., \& {Ayres}, T.~R. 2003, \apj, 593, 534,
  \dodoi{10.1086/376405}

\bibitem[{{Kastner} {et~al.}(2002){Kastner}, {Huenemoerder}, {Schulz},
  {Canizares}, \& {Weintraub}}]{Kastner+02}
{Kastner}, J.~H., {Huenemoerder}, D.~P., {Schulz}, N.~S., {Canizares}, C.~R.,
  \& {Weintraub}, D.~A. 2002, \apj, 567, 434, \dodoi{10.1086/338419}

\bibitem[{{Kastner} {et~al.}(2004){Kastner}, {Richmond}, {Grosso}, {Weintraub},
  {Simon}, {Frank}, {Hamaguchi}, {Ozawa}, \& {Henden}}]{Kastner+04}
{Kastner}, J.~H., {Richmond}, M., {Grosso}, N., {et~al.} 2004, \nat, 430, 429,
  \dodoi{10.1038/nature02747}

\bibitem[{{Kennedy} \& {Kenyon}(2009)}]{Kennedy+Kenyon09}
{Kennedy}, G.~M., \& {Kenyon}, S.~J. 2009, \apj, 695, 1210,
  \dodoi{10.1088/0004-637X/695/2/1210}

\bibitem[{{Kimura} {et~al.}(2016){Kimura}, {Kunitomo}, \&
  {Takahashi}}]{Kimura+16}
{Kimura}, S.~S., {Kunitomo}, M., \& {Takahashi}, S.~Z. 2016, \mnras, 461, 2257,
  \dodoi{10.1093/mnras/stw1531}

\bibitem[{{Kippenhahn} \& {Weigert}(1990)}]{Kippenhahn+Weigert90}
{Kippenhahn}, R., \& {Weigert}, A. 1990, {Stellar Structure and Evolution}
  (Springer-Verlag)

\bibitem[{{Kobayashi} \& {Tanaka}(2010)}]{Kobayashi+Tanaka10}
{Kobayashi}, H., \& {Tanaka}, H. 2010, \icarus, 206, 735,
  \dodoi{10.1016/j.icarus.2009.10.004}

\bibitem[{{Komaki} {et~al.}(2020){Komaki}, {Nakatani}, \&
  {Yoshida}}]{Komaki+20}
{Komaki}, A., {Nakatani}, R., \& {Yoshida}, N. 2020, arXiv e-prints,
  arXiv:2012.14852.
\newblock \doarXiv{2012.14852}

\bibitem[{{Kuffmeier} {et~al.}(2018){Kuffmeier}, {Frimann}, {Jensen}, \&
  {Haugb{\o}lle}}]{Kuffmeier+18}
{Kuffmeier}, M., {Frimann}, S., {Jensen}, S.~S., \& {Haugb{\o}lle}, T. 2018,
  \mnras, 475, 2642, \dodoi{10.1093/mnras/sty024}

\bibitem[{{Kunitomo} {et~al.}(2017){Kunitomo}, {Guillot}, {Takeuchi}, \&
  {Ida}}]{Kunitomo+17}
{Kunitomo}, M., {Guillot}, T., {Takeuchi}, T., \& {Ida}, S. 2017, \aap, 599,
  A49, \dodoi{10.1051/0004-6361/201628260}

\bibitem[{{Kunitomo} {et~al.}(2011){Kunitomo}, {Ikoma}, {Sato}, {Katsuta}, \&
  {Ida}}]{Kunitomo+11}
{Kunitomo}, M., {Ikoma}, M., {Sato}, B., {Katsuta}, Y., \& {Ida}, S. 2011,
  \apj, 737, 66, \dodoi{10.1088/0004-637X/737/2/66}

\bibitem[{{Kunitomo} {et~al.}(2020){Kunitomo}, {Suzuki}, \&
  {Inutsuka}}]{Kunitomo+20}
{Kunitomo}, M., {Suzuki}, T.~K., \& {Inutsuka}, S.-i. 2020, \mnras, 492, 3849,
  \dodoi{10.1093/mnras/staa087}

\bibitem[{{Lesur} {et~al.}(2013){Lesur}, {Ferreira}, \& {Ogilvie}}]{Lesur+13}
{Lesur}, G., {Ferreira}, J., \& {Ogilvie}, G.~I. 2013, \aap, 550, A61,
  \dodoi{10.1051/0004-6361/201220395}

\bibitem[{{Liffman}(2003)}]{Liffman03}
{Liffman}, K. 2003, \pasa, 20, 337, \dodoi{10.1071/AS03019}

\bibitem[{{Long} {et~al.}(2019){Long}, {Herczeg}, {Harsono}, {Pinilla},
  {Tazzari}, {Manara}, {Pascucci}, {Cabrit}, {Nisini}, {Johnstone}, {Edwards},
  {Salyk}, {Menard}, {Lodato}, {Boehler}, {Mace}, {Liu}, {Mulders}, {Hendler},
  {Ragusa}, {Fischer}, {Banzatti}, {Rigliaco}, {van de Plas}, {Dipierro},
  {Gully-Santiago}, \& {Lopez-Valdivia}}]{Long+19}
{Long}, F., {Herczeg}, G.~J., {Harsono}, D., {et~al.} 2019, \apj, 882, 49,
  \dodoi{10.3847/1538-4357/ab2d2d}

\bibitem[{{Lynden-Bell} \& {Pringle}(1974)}]{Lynden-Bell+Pringle74}
{Lynden-Bell}, D., \& {Pringle}, J.~E. 1974, \mnras, 168, 603

\bibitem[{{Mangeney} \& {Praderie}(1984)}]{Mangeney+Praderie84}
{Mangeney}, A., \& {Praderie}, F. 1984, \aap, 130, 143

\bibitem[{{Miley} {et~al.}(2021){Miley}, {Pani{\'c}}, {Booth}, {Ilee}, {Ida},
  \& {Kunitomo}}]{Miley+21}
{Miley}, J.~M., {Pani{\'c}}, O., {Booth}, R.~A., {et~al.} 2021, \mnras, 500,
  4658, \dodoi{10.1093/mnras/staa3517}

\bibitem[{{Miley} {et~al.}(2019){Miley}, {Pani{\'c}}, {Haworth}, {Pascucci},
  {Wyatt}, {Clarke}, {Richards}, \& {Ratzka}}]{Miley+19}
{Miley}, J.~M., {Pani{\'c}}, O., {Haworth}, T.~J., {et~al.} 2019, \mnras, 485,
  739, \dodoi{10.1093/mnras/stz426}

\bibitem[{{Miyake} {et~al.}(2016){Miyake}, {Suzuki}, \& {Inutsuka}}]{Miyake+16}
{Miyake}, T., {Suzuki}, T.~K., \& {Inutsuka}, S.-i. 2016, \apj, 821, 3,
  \dodoi{10.3847/0004-637X/821/1/3}

\bibitem[{{Mohanty} {et~al.}(2013){Mohanty}, {Greaves}, {Mortlock}, {Pascucci},
  {Scholz}, {Thompson}, {Apai}, {Lodato}, \& {Looper}}]{Mohanty+13}
{Mohanty}, S., {Greaves}, J., {Mortlock}, D., {et~al.} 2013, \apj, 773, 168,
  \dodoi{10.1088/0004-637X/773/2/168}

\bibitem[{{Muro-Arena} {et~al.}(2020){Muro-Arena}, {Benisty}, {Ginski},
  {Dominik}, {Facchini}, {Villenave}, {van Boekel}, {Chauvin}, {Garufi},
  {Henning}, {Janson}, {Keppler}, {Matter}, {M{\'e}nard}, {Stolker}, {Zurlo},
  {Blanchard}, {Maurel}, {Moeller-Nilsson}, {Petit}, {Roux}, {Sevin}, \&
  {Wildi}}]{Muro-Arena+20}
{Muro-Arena}, G.~A., {Benisty}, M., {Ginski}, C., {et~al.} 2020, \aap, 635,
  A121, \dodoi{10.1051/0004-6361/201936509}

\bibitem[{{Muzerolle} {et~al.}(2005){Muzerolle}, {Luhman}, {Brice{\~n}o},
  {Hartmann}, \& {Calvet}}]{Muzerolle+05}
{Muzerolle}, J., {Luhman}, K.~L., {Brice{\~n}o}, C., {Hartmann}, L., \&
  {Calvet}, N. 2005, \apj, 625, 906, \dodoi{10.1086/429483}

\bibitem[{{Nakatani} {et~al.}(2018{\natexlab{a}}){Nakatani}, {Hosokawa},
  {Yoshida}, {Nomura}, \& {Kuiper}}]{Nakatani+18b}
{Nakatani}, R., {Hosokawa}, T., {Yoshida}, N., {Nomura}, H., \& {Kuiper}, R.
  2018{\natexlab{a}}, \apj, 865, 75, \dodoi{10.3847/1538-4357/aad9fd}

\bibitem[{{Nakatani} {et~al.}(2018{\natexlab{b}}){Nakatani}, {Hosokawa},
  {Yoshida}, {Nomura}, \& {Kuiper}}]{Nakatani+18a}
---. 2018{\natexlab{b}}, \apj, 857, 57, \dodoi{10.3847/1538-4357/aab70b}

\bibitem[{{Nakatani} {et~al.}(2020){Nakatani}, {Kobayashi}, {Kuiper}, {Nomura},
  \& {Aikawa}}]{Nakatani+20}
{Nakatani}, R., {Kobayashi}, H., {Kuiper}, R., {Nomura}, H., \& {Aikawa}, Y.
  2020, arXiv e-prints, arXiv:2009.06438.
\newblock \doarXiv{2009.06438}

\bibitem[{{Noyes} {et~al.}(1984){Noyes}, {Hartmann}, {Baliunas}, {Duncan}, \&
  {Vaughan}}]{Noyes+84}
{Noyes}, R.~W., {Hartmann}, L.~W., {Baliunas}, S.~L., {Duncan}, D.~K., \&
  {Vaughan}, A.~H. 1984, \apj, 279, 763, \dodoi{10.1086/161945}

\bibitem[{{Ogihara} {et~al.}(2020){Ogihara}, {Kunitomo}, \&
  {Hori}}]{Ogihara+20}
{Ogihara}, M., {Kunitomo}, M., \& {Hori}, Y. 2020, \apj, 899, 91,
  \dodoi{10.3847/1538-4357/aba75e}

\bibitem[{{Owen} {et~al.}(2012){Owen}, {Clarke}, \& {Ercolano}}]{Owen+12}
{Owen}, J.~E., {Clarke}, C.~J., \& {Ercolano}, B. 2012, \mnras, 422, 1880,
  \dodoi{10.1111/j.1365-2966.2011.20337.x}

\bibitem[{{Owen} {et~al.}(2010){Owen}, {Ercolano}, {Clarke}, \&
  {Alexander}}]{Owen+10}
{Owen}, J.~E., {Ercolano}, B., {Clarke}, C.~J., \& {Alexander}, R.~D. 2010,
  \mnras, 401, 1415, \dodoi{10.1111/j.1365-2966.2009.15771.x}

\bibitem[{{Owen} \& {Kollmeier}(2019)}]{Owen+Kollmeier19}
{Owen}, J.~E., \& {Kollmeier}, J.~A. 2019, \mnras, 487, 3702,
  \dodoi{10.1093/mnras/stz1591}

\bibitem[{{Pani{\'c}} {et~al.}(2008){Pani{\'c}}, {Hogerheijde}, {Wilner}, \&
  {Qi}}]{Panic+08}
{Pani{\'c}}, O., {Hogerheijde}, M.~R., {Wilner}, D., \& {Qi}, C. 2008, \aap,
  491, 219, \dodoi{10.1051/0004-6361:20079261}

\bibitem[{{Parravano} {et~al.}(2003){Parravano}, {Hollenbach}, \&
  {McKee}}]{Parravano+03}
{Parravano}, A., {Hollenbach}, D.~J., \& {McKee}, C.~F. 2003, \apj, 584, 797,
  \dodoi{10.1086/345807}

\bibitem[{{Pascucci} {et~al.}(2016){Pascucci}, {Testi}, {Herczeg}, {Long},
  {Manara}, {Hendler}, {Mulders}, {Krijt}, {Ciesla}, {Henning}, {Mohanty},
  {Drabek-Maunder}, {Apai}, {Sz{\H u}cs}, {Sacco}, \& {Olofsson}}]{Pascucci+16}
{Pascucci}, I., {Testi}, L., {Herczeg}, G.~J., {et~al.} 2016, \apj, 831, 125,
  \dodoi{10.3847/0004-637X/831/2/125}

\bibitem[{{Paxton} {et~al.}(2011){Paxton}, {Bildsten}, {Dotter}, {Herwig},
  {Lesaffre}, \& {Timmes}}]{Paxton+11}
{Paxton}, B., {Bildsten}, L., {Dotter}, A., {et~al.} 2011, \apjs, 192, 3,
  \dodoi{10.1088/0067-0049/192/1/3}

\bibitem[{{Picogna} {et~al.}(2019){Picogna}, {Ercolano}, {Owen}, \&
  {Weber}}]{Picogna+19}
{Picogna}, G., {Ercolano}, B., {Owen}, J.~E., \& {Weber}, M.~L. 2019, \mnras,
  487, 691, \dodoi{10.1093/mnras/stz1166}

\bibitem[{{Pinte} {et~al.}(2016){Pinte}, {Dent}, {M{\'e}nard}, {Hales}, {Hill},
  {Cortes}, \& {de Gregorio-Monsalvo}}]{Pinte+16}
{Pinte}, C., {Dent}, W.~R.~F., {M{\'e}nard}, F., {et~al.} 2016, \apj, 816, 25,
  \dodoi{10.3847/0004-637X/816/1/25}

\bibitem[{{Preibisch} {et~al.}(2005){Preibisch}, {Kim}, {Favata}, {Feigelson},
  {Flaccomio}, {Getman}, {Micela}, {Sciortino}, {Stassun}, {Stelzer}, \&
  {Zinnecker}}]{Preibisch+05}
{Preibisch}, T., {Kim}, Y.-C., {Favata}, F., {et~al.} 2005, \apjs, 160, 401,
  \dodoi{10.1086/432891}

\bibitem[{{Rasio} {et~al.}(1996){Rasio}, {Tout}, {Lubow}, \&
  {Livio}}]{Rasio+96}
{Rasio}, F.~A., {Tout}, C.~A., {Lubow}, S.~H., \& {Livio}, M. 1996, \apj, 470,
  1187, \dodoi{10.1086/177941}

\bibitem[{{Rebull} {et~al.}(2004){Rebull}, {Wolff}, \& {Strom}}]{Rebull+04}
{Rebull}, L.~M., {Wolff}, S.~C., \& {Strom}, S.~E. 2004, \aj, 127, 1029,
  \dodoi{10.1086/380931}

\bibitem[{{Reffert} {et~al.}(2015){Reffert}, {Bergmann}, {Quirrenbach},
  {Trifonov}, \& {K{\"u}nstler}}]{Reffert+15}
{Reffert}, S., {Bergmann}, C., {Quirrenbach}, A., {Trifonov}, T., \&
  {K{\"u}nstler}, A. 2015, \aap, 574, A116, \dodoi{10.1051/0004-6361/201322360}

\bibitem[{{Ribas} {et~al.}(2015){Ribas}, {Bouy}, \& {Mer{\'{\i}}n}}]{Ribas+15}
{Ribas}, {\'A}., {Bouy}, H., \& {Mer{\'{\i}}n}, B. 2015, \aap, 576, A52,
  \dodoi{10.1051/0004-6361/201424846}

\bibitem[{{Rodenkirch} {et~al.}(2020){Rodenkirch}, {Klahr}, {Fendt}, \&
  {Dullemond}}]{Rodenkirch+20}
{Rodenkirch}, P.~J., {Klahr}, H., {Fendt}, C., \& {Dullemond}, C.~P. 2020,
  \aap, 633, A21, \dodoi{10.1051/0004-6361/201834945}

\bibitem[{{Sato} {et~al.}(2008){Sato}, {Izumiura}, {Toyota}, {Kambe}, {Ikoma},
  {Omiya}, {Masuda}, {Takeda}, {Murata}, {Itoh}, {Ando}, {Yoshida}, {Kokubo},
  \& {Ida}}]{Sato+08}
{Sato}, B., {Izumiura}, H., {Toyota}, E., {et~al.} 2008, \pasj, 60, 539,
  \dodoi{10.1093/pasj/60.3.539}

\bibitem[{{Serenelli} {et~al.}(2009){Serenelli}, {Basu}, {Ferguson}, \&
  {Asplund}}]{Serenelli+09}
{Serenelli}, A.~M., {Basu}, S., {Ferguson}, J.~W., \& {Asplund}, M. 2009,
  \apjl, 705, L123, \dodoi{10.1088/0004-637X/705/2/L123}

\bibitem[{{Shakura} \& {Sunyaev}(1973)}]{Shakura+Sunyaev73}
{Shakura}, N.~I., \& {Sunyaev}, R.~A. 1973, \aap, 24, 337

\bibitem[{{Siess} {et~al.}(2000){Siess}, {Dufour}, \& {Forestini}}]{Siess+00}
{Siess}, L., {Dufour}, E., \& {Forestini}, M. 2000, \aap, 358, 593

\bibitem[{{Stahler}(1988)}]{Stahler88}
{Stahler}, S.~W. 1988, \apj, 332, 804, \dodoi{10.1086/166694}

\bibitem[{{Stahler} \& {Palla}(2004)}]{Stahler+Palla04}
{Stahler}, S.~W., \& {Palla}, F. 2004, {The Formation of Stars} (Weinheim:
  Wiley-VCH)

\bibitem[{{Stelzer} {et~al.}(2009){Stelzer}, {Robrade}, {Schmitt}, \&
  {Bouvier}}]{Stelzer+09}
{Stelzer}, B., {Robrade}, J., {Schmitt}, J.~H.~M.~M., \& {Bouvier}, J. 2009,
  \aap, 493, 1109, \dodoi{10.1051/0004-6361:200810540}

\bibitem[{{Suzuki} {et~al.}(2013){Suzuki}, {Imada}, {Kataoka}, {Kato},
  {Matsumoto}, {Miyahara}, \& {Tsuneta}}]{Suzuki+13}
{Suzuki}, T.~K., {Imada}, S., {Kataoka}, R., {et~al.} 2013, \pasj, 65, 98,
  \dodoi{10.1093/pasj/65.5.98}

\bibitem[{{Suzuki} \& {Inutsuka}(2009)}]{Suzuki+Inutsuka09}
{Suzuki}, T.~K., \& {Inutsuka}, S.-i. 2009, \apjl, 691, L49,
  \dodoi{10.1088/0004-637X/691/1/L49}

\bibitem[{{Suzuki} {et~al.}(2016){Suzuki}, {Ogihara}, {Morbidelli}, {Crida}, \&
  {Guillot}}]{Suzuki+16}
{Suzuki}, T.~K., {Ogihara}, M., {Morbidelli}, A., {Crida}, A., \& {Guillot}, T.
  2016, \aap, 596, A74, \dodoi{10.1051/0004-6361/201628955}

\bibitem[{{Takahashi} {et~al.}(2013){Takahashi}, {Inutsuka}, \&
  {Machida}}]{Takahashi+13}
{Takahashi}, S.~Z., {Inutsuka}, S.-i., \& {Machida}, M.~N. 2013, \apj, 770, 71,
  \dodoi{10.1088/0004-637X/770/1/71}

\bibitem[{{Takasao} {et~al.}(2018){Takasao}, {Tomida}, {Iwasaki}, \&
  {Suzuki}}]{Takasao+18}
{Takasao}, S., {Tomida}, K., {Iwasaki}, K., \& {Suzuki}, T.~K. 2018, \apj, 857,
  4, \dodoi{10.3847/1538-4357/aab5b3}

\bibitem[{{Takeuchi} {et~al.}(2005){Takeuchi}, {Clarke}, \&
  {Lin}}]{Takeuchi+05}
{Takeuchi}, T., {Clarke}, C.~J., \& {Lin}, D.~N.~C. 2005, \apj, 627, 286,
  \dodoi{10.1086/430393}

\bibitem[{{Taki} {et~al.}(2016){Taki}, {Fujimoto}, \& {Ida}}]{Taki+16}
{Taki}, T., {Fujimoto}, M., \& {Ida}, S. 2016, \aap, 591, A86,
  \dodoi{10.1051/0004-6361/201527732}

\bibitem[{{Taki} {et~al.}(2020){Taki}, {Kuwabara}, {Kobayashi}, \&
  {Suzuki}}]{Taki+20}
{Taki}, T., {Kuwabara}, K., {Kobayashi}, H., \& {Suzuki}, T.~K. 2020, arXiv
  e-prints, arXiv:2004.08839.
\newblock \doarXiv{2004.08839}

\bibitem[{{Tanaka} {et~al.}(2013){Tanaka}, {Nakamoto}, \& {Omukai}}]{Tanaka+13}
{Tanaka}, K.~E.~I., {Nakamoto}, T., \& {Omukai}, K. 2013, \apj, 773, 155,
  \dodoi{10.1088/0004-637X/773/2/155}

\bibitem[{{Telleschi} {et~al.}(2007){Telleschi}, {G{\"u}del}, {Briggs},
  {Audard}, \& {Palla}}]{Telleschi+07a}
{Telleschi}, A., {G{\"u}del}, M., {Briggs}, K.~R., {Audard}, M., \& {Palla}, F.
  2007, \aap, 468, 425, \dodoi{10.1051/0004-6361:20066565}

\bibitem[{{Tognelli} {et~al.}(2015){Tognelli}, {Prada Moroni}, \&
  {Degl'Innocenti}}]{Tognelli+15}
{Tognelli}, E., {Prada Moroni}, P.~G., \& {Degl'Innocenti}, S. 2015, \mnras,
  454, 4037, \dodoi{10.1093/mnras/stv2254}

\bibitem[{{Tu} {et~al.}(2015){Tu}, {Johnstone}, {G{\"u}del}, \&
  {Lammer}}]{Tu+15}
{Tu}, L., {Johnstone}, C.~P., {G{\"u}del}, M., \& {Lammer}, H. 2015, \aap, 577,
  L3, \dodoi{10.1051/0004-6361/201526146}

\bibitem[{{Turner} {et~al.}(2014){Turner}, {Fromang}, {Gammie}, {Klahr},
  {Lesur}, {Wardle}, \& {Bai}}]{Turner+14}
{Turner}, N.~J., {Fromang}, S., {Gammie}, C., {et~al.} 2014, {Transport and
  Accretion in Planet-Forming Disks} (University of Arizona Press), 411--432,
  \dodoi{10.2458/azu_uapress_9780816531240-ch018}

\bibitem[{{van den Ancker} {et~al.}(1997){van den Ancker}, {The}, {Tjin A
  Djie}, {Catala}, {de Winter}, {Blondel}, \& {Waters}}]{vandenAncker97}
{van den Ancker}, M.~E., {The}, P.~S., {Tjin A Djie}, H.~R.~E., {et~al.} 1997,
  \aap, 324, L33

\bibitem[{{van der Walt} {et~al.}(2011){van der Walt}, {Colbert}, \&
  {Varoquaux}}]{vanderWalt+11}
{van der Walt}, S., {Colbert}, S.~C., \& {Varoquaux}, G. 2011, Computing in
  Science and Engineering, 13, 22, \dodoi{10.1109/MCSE.2011.37}

\bibitem[{Velikhov(1959)}]{Velikhov59}
Velikhov, E. 1959, Sov. Phys. JETP, 36, 1398

\bibitem[{{Vilhu} \& {Rucinski}(1983)}]{Vilhu+Rucinski83}
{Vilhu}, O., \& {Rucinski}, S.~M. 1983, \aap, 127, 5

\bibitem[{{Villaver} \& {Livio}(2009)}]{Villaver+Livio09}
{Villaver}, E., \& {Livio}, M. 2009, \apjl, 705, L81,
  \dodoi{10.1088/0004-637X/705/1/L81}

\bibitem[{{Villebrun} {et~al.}(2019){Villebrun}, {Alecian}, {Hussain},
  {Bouvier}, {Folsom}, {Lebreton}, {Amard}, {Charbonnel}, {Gallet},
  {Haemmerl{\'e}}, {B{\"o}hm}, {Johns-Krull}, {Kochukhov}, {Marsden}, {Morin},
  \& {Petit}}]{Villebrun+19}
{Villebrun}, F., {Alecian}, E., {Hussain}, G., {et~al.} 2019, \aap, 622, A72,
  \dodoi{10.1051/0004-6361/201833545}

\bibitem[{{Wang} {et~al.}(2019){Wang}, {Bai}, \& {Goodman}}]{Wang+19}
{Wang}, L., {Bai}, X.-N., \& {Goodman}, J. 2019, \apj, 874, 90,
  \dodoi{10.3847/1538-4357/ab06fd}

\bibitem[{{Wang} \& {Goodman}(2017)}]{Wang+Goodman17}
{Wang}, L., \& {Goodman}, J. 2017, \apj, 847, 11,
  \dodoi{10.3847/1538-4357/aa8726}

\bibitem[{{Williams} \& {Cieza}(2011)}]{Williams+Cieza11}
{Williams}, J.~P., \& {Cieza}, L.~A. 2011, \araa, 49, 67,
  \dodoi{10.1146/annurev-astro-081710-102548}

\bibitem[{{Wright} {et~al.}(2011){Wright}, {Drake}, {Mamajek}, \&
  {Henry}}]{Wright+11}
{Wright}, N.~J., {Drake}, J.~J., {Mamajek}, E.~E., \& {Henry}, G.~W. 2011,
  \apj, 743, 48, \dodoi{10.1088/0004-637X/743/1/48}

\bibitem[{{Yasui} {et~al.}(2014){Yasui}, {Kobayashi}, {Tokunaga}, \&
  {Saito}}]{Yasui+14}
{Yasui}, C., {Kobayashi}, N., {Tokunaga}, A.~T., \& {Saito}, M. 2014, \mnras,
  442, 2543, \dodoi{10.1093/mnras/stu1013}

\bibitem[{{Zahn}(1977)}]{Zahn+77}
{Zahn}, J.-P. 1977, \aap, 57, 383

\bibitem[{{Zinnecker} \& {Preibisch}(1994)}]{Zinnecker+Preibisch94}
{Zinnecker}, H., \& {Preibisch}, T. 1994, \aap, 292, 152

\end{thebibliography}
\bibliographystyle{aasjournal}



\end{document}